\documentstyle[12pt,epsf,psfig]{article}

\newcommand{\gtrsim}{ \mathop{}_{\textstyle \sim}^{\textstyle >} }
\newcommand{\lesssim}{ \mathop{}_{\textstyle \sim}^{\textstyle <} }

\def\thefootnote{\fnsymbol{footnote}}

\begin{document}

\begin{titlepage}
\begin{center}

\hfill    LBNL-42005\\
\hfill    hep-ph/9807265\\
\hfill    July, 1998\\

\vskip .5in

{\large \bf Cosmological Moduli Problem in a Supersymmetric Model\\
with Direct Gauge Mediation}\footnote
 {This work was supported by the Director, Office of Energy Research,
Office of Basic Energy Services, of the U.S.~Department of Energy
under Contract DE-AC03-76SF00098.}

\vskip .50in

{\large Takeo~Moroi}

\vskip .5in

{\it Theoretical Physics Group, Lawrence Berkeley National
Laboratory\\ University of California, Berkeley, CA 94720, U.S.A.}

\end{center}

\vskip .5in

\begin{abstract}

Recently, an interesting class of the direct gauge mediation
supersymmetry (SUSY) breaking models are proposed, in which the
minimum of the potential of the SUSY breaking field is determined by
the inverted hierarchy mechanism. We consider their cosmological
implications. In this class of models, SUSY breaking field has a very
flat potential, which may have a cosmological importance. Assuming the
initial amplitude of the SUSY breaking field to be of the order of the
Planck scale, it can be a source of a large entropy production.  A
special attention is paid to the cosmological moduli problem, and we
will see the cosmological mass density of the moduli field can be
significantly reduced.

\end{abstract}
\end{titlepage}

\renewcommand{\thepage}{\arabic{page}}
\setcounter{page}{1}
\renewcommand{\thefootnote}{\#\arabic{footnote}}
\setcounter{footnote}{0}

\section{Introduction}
 \label{sec:intro}
 \setcounter{equation}{0}

Low energy supersymmetry (SUSY) has been regarded as one of the
attractive new physics beyond the standard model, since it may provide
a natural explanation to the stability of the electroweak scale
against radiative corrections. However, contrary to our theoretical
interests, any superpartner of the standard model particle has not
discovered yet, and hence SUSY has to be broken in nature.
Unfortunately, we do not have a clear picture of the SUSY breaking in
nature, and the understanding of the origin of the SUSY breaking is
one of the most important issues in the study of the supersymmetric
models.

In recent years, a new framework for the SUSY breaking, i.e., gauge
mediated SUSY breaking (GMSB)~\cite{GMSB_origs}, has been attracting
many interests, and its phenomenological implications have been
extensively investigated~\cite{hph9801271}. In particular, in GMSB,
SUSY breaking in the SUSY standard model (SSM) sector is mediated by
the gauge interactions which do not distinguish flavors.  In this
scheme, dangerous off-diagonal elements in the sfermion mass matrices
are suppressed, and serious SUSY flavor changing neutral current
(FCNC) problem can be evaded.

In spite of the phenomenological interests, however, cosmology of GMSB
is not fully satisfactory. In particular, relic abundances of the
gravitino~\cite{PRL48-223,PLB303-289} and the moduli
field~\cite{PRD56-1281} have been known to be problematic. These
problems are often called ``gravitino problem'' and ``cosmological
moduli problem.'' In some sense, they are more serious than the
gravity mediated SUSY breaking case, and they may be crucial weak
points of GMSB. (These issues will be reviewed in the next section.)
However, one should note that there are rooms to solve or improve
these difficulties; since these difficulties are based on a kind of
``minimal'' assumption, some of them may be solved or relaxed by a new
idea. For example, thermal inflation~\cite{thermal_inf} is proposed to
dilute the unwanted particles.

In this paper, we would like to propose a new mechanism for a large
entropy production, which can be a resolution to the cosmological
problems in GMSB. Our scenario is based on a class of models with
direct gauge mediation in which the messenger particles have a direct
coupling to the original SUSY breaking field. In particular, recently,
several direct gauge mediation models are proposed in which SUSY
breaking field has exactly flat potential at the tree
level~\cite{PRL79-18,NPB510-12,PLB413-336,hph9804450}. (For other
classes of models of direct gauge mediation, see
Refs.~\cite{nonpert,otherDGM}.)  In this class of models, minimum of
the potential of the SUSY breaking field is determined by the inverted
hierarchy mechanism~\cite{PLB105-267}, and SUSY breaking field has a
very flat potential even after the potential is lifted. In this case,
SUSY breaking field may play an important role to dilute unwanted
particles; with an assumption that the SUSY breaking field has an
initial amplitude of the order of the Planck scale, various
cosmological problems can be naturally solved. One virtue of this
scenario is that the source of the large entropy production is already
in the framework of the SUSY breaking mechanism. Therefore, the
scenario is fairly economical, and we do not have to introduce any new
field only for the entropy production (like ``flaton''), contrary to
the case of thermal inflation~\cite{thermal_inf}. In the following
sections, we see how this works, and consider if we may have a
cosmologically consistent scenario.

The organization of this paper is as follows.  In
Section~\ref{sec:overview}, we give an overview of the cosmological
difficulties in GMSB. Then, in Section~\ref{sec:model}, we briefly
review the important aspects of the direct gauge mediation model with
the inverted hierarchy mechanism. In particular, the SUSY breaking
field plays a very important role in our discussion, so we see the
properties of the potential of the SUSY breaking field in some
detail. In Section~\ref{sec:evolution}, cosmology based on the direct
gauge mediation model with the inverted hierarchy mechanism is
discussed. In particular, we concentrate on the cosmological evolution
of the SUSY breaking field, and we discuss how it improves the
cosmological difficulties.  Section~\ref{sec:discussion} is devoted to
discussion.

\section{Overview of the Cosmology of GMSB}
 \label{sec:overview}
 \setcounter{equation}{0}

Before discussing the cosmology of the direct gauge mediation model,
let us first briefly overview the cosmology of the gauge mediation
model.

Cosmologically, one important outcome of the GMSB is the light
gravitino; the gravitino mass $m_{3/2}$ in this scheme has to be much
lighter than the SSM scale $m_{\rm SSM}$. This is because mass squared
matrices of sfermions have off-diagonal elements of $O(m_{3/2}^2)$. If
this effect is comparable to the dominant contribution from GMSB,
dangerous SUSY FCNC problem arises again, which spoils the important
motivation of GMSB.  Consequently, the gravitino becomes the lightest
superparticle (LSP) in this scenario.

Keeping this feature in mind, one of the most famous cosmological
constraint in GMSB is from the mass density of the gravitino in the
Universe. If the gravitino is thermalized in the early Universe, and
if it is not diluted, its mass density may significantly contribute to
the energy density of the Universe. Without dilution, mass density of
the gravitino is proportional to $m_{3/2}$, and the Universe is
overclosed if the gravitino mass is heavier than about
1~keV~\cite{PRL48-223}.

\begin{figure}
 \centerline{\epsfxsize=0.5\textwidth\epsfbox{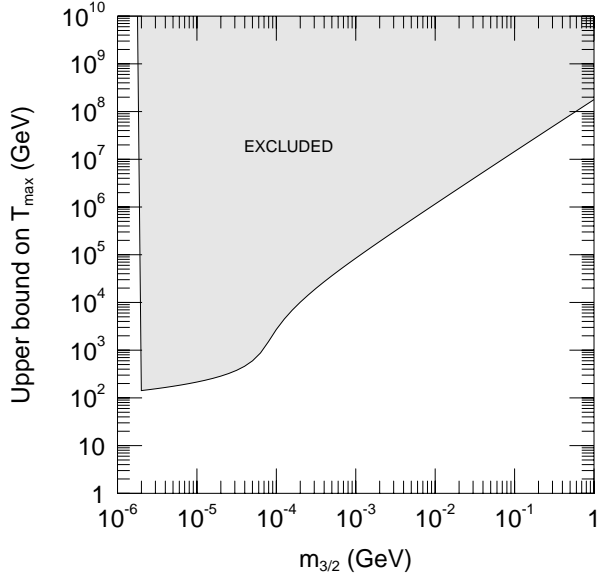}}
 \caption{Upper bound on the maximal temperature of the Universe
as a function of the gravitino mass $m_{3/2}$.}
 \label{fig:rho_grvtno}
 \end{figure}

With an enough entropy production after the decoupling of the
gravitino from the thermal bath, gravitino mass heavier than 1~keV may
be also viable.  However, even in this case, the gravitinos are
produced in the thermal bath due to scattering and decay processes.
These secondary gravitinos also contribute to the mass density of the
Universe, and hence gravitino production after the entropy production
has to be inefficient. Since the gravitino production is more
effective for higher temperature, we obtain an upper bound on the
maximal temperature of the Universe after the late entropy production
in order not to overclose the Universe~\cite{PLB303-289}. Notice that
the interaction of the longitudinal gravitino with matter is
proportional to $m_{3/2}^{-1}$, and hence the constraint becomes
weaker for a heavier gravitino. In Fig.~\ref{fig:rho_grvtno}, we show
the upper bound on the maximal temperature $T_{\rm max}$ as a function
of the gravitino mass; the upper bound is from $\sim 100{\rm ~GeV}$ to
$\sim 10^8{\rm ~GeV}$, for the gravitino mass 1~keV -- 1~GeV. In
particular, the constraint is very strict in the case of ${\rm
1~keV}\lesssim m_{3/2}\lesssim{\rm 100~keV}$, where the gravitinos are
mainly produced by the decay processes.  If the Universe starts with a
temperature higher than this upper bound, the Universe is overclosed
by the gravitino, unless there is an enough entropy production below
this temperature.

More serious problem arises in the framework of superstring models. In
superstring models, dilaton and moduli fields (which we call
``moduli'' fields hereafter) exist which are the flat directions
related to symmetries in the superstring theory. Mass of the moduli
field originates to SUSY breaking effect, and is of the same order of
the gravitino mass. (Throughout this paper, we approximate the mass of
the moduli field to be the gravitino mass $m_{3/2}$.) Generically,
moduli field takes an initial amplitude of the order of the Planck
scale, unless our vacuum lies at or near a point of enhanced
symmetry~\cite{DinRanTho}.\footnote
 {However, it is difficult to construct a realistic model which has
our vacuum as a symmetry enhanced point.}
 Then, it starts oscillation when the expansion rate of the Universe
becomes comparable to the mass of the moduli field. Since the
interactions of the moduli field are suppressed by inverse powers of
the Planck scale, moduli field lighter than about 100~MeV has a
lifetime longer than the present age of the Universe. In GMSB, this is
(almost) always the case. In this case, mass density of the moduli
field becomes enormous, if there is no dilution.  Assuming the
radiation dominance before the moduli field starts to move, naive
calculation results in the density parameter of the moduli field as
 \begin{eqnarray}
  \Omega_\phi h^2 \sim 6\times 10^{14} \times
  \left(\frac{g_*}{100}\right)^{-1/4}
  \left(\frac{m_{3/2}}{\rm 100~keV}\right)^{1/2}
  \left(\frac{\phi_0}{M_*}\right)^2,
 \label{Omg_noD}
 \end{eqnarray}
 where $h$ is the Hubble constant in units of 100~km/sec/Mpc, $g_*$ is
the effective number of the massless degrees of freedom when the
moduli field starts to move, $M_*\simeq 2.4\times 10^{18}{\rm ~GeV}$
is the reduced Planck scale, and $\phi_0$ is the initial amplitude of
the moduli field. For example, for $m_{3/2}={\rm 100~keV}$, the
initial amplitude has to be smaller than $\sim 10^{-7}M_*$ in order
not to overclose the Universe, and this is an extreme fine tuning of
the initial condition. In other words, if the initial amplitude takes
the natural value (i.e., $\sim M_*$), a large entropy production is
inevitable for a viable cosmological scenario.

Even if we adopt a large entropy production to dilute the gravitino
and the moduli field away, it is still non-trivial whether a
consistent cosmological scenario can be obtained. If there is a large
entropy production, it also dilutes the possible baryon asymmetry
generated in the early stage of the Universe. Naively speaking, baryon
number asymmetry has to be generated after the entropy
production. However, in some case, the reheating temperature after the
entropy production becomes too low for baryogenesis. In
Ref.~\cite{PRD56-1281}, it has been pointed out that Affleck-Dine
mechanism for baryogenesis~\cite{NPB249-361} may be able to generate
enough baryon asymmetry even if there is a large entropy production.
This topics is reviewed later.  Candidates of the cold dark matter
(CDM) is another interesting issue in GMSB.  In the gravity mediated
SUSY breaking scenario, the lightest neutralino~\cite{PRep267-195} or
sneutrino~\cite{hph9712515} can be a promising candidate of the CDM,
if it is the LSP. However, in GMSB, they cannot be the CDM, since they
can decay into gravitino and their superpartner. Several candidates of
the CDM are proposed in the framework of
GMSB~\cite{PLB386-189,messenger-DM}, but these candidates are also
diluted by the entropy production. One interesting candidate is the
coherent oscillation of the moduli field, if its energy density is
diluted enough.  Therefore, it is very important to consider the
possibility to dilute the energy density of the moduli field down to
$\Omega_\phi\lesssim 1$.

Another class of cosmological problems are related to the structure of
the scalar potential; the scalar potential may have unwanted minimum
which is deeper than the phenomenologically viable local minimum. For
example, original low energy gauge mediation model may have a color
breaking minimum~\cite{QCD_br}, and the models proposed in
Refs.~\cite{NPB510-12,PLB413-336,hph9804450} have a SUSY preserving
true vacuum at the origin of the potential of the SUSY breaking
field.\footnote
 {However, the structure of the potential is model-dependent, and
models without unwanted minimum may be constructed.}
 Usually, the tunneling rate to the true vacuum can be so small that
the transition does not happen for the time scale of the age of the
Universe. Therefore, it is phenomenologically consistent once the SUSY
breaking field is trapped in the minimum we want. However,
cosmologically, we have not understood how the SUSY breaking field is
trapped in the relevant (local) minimum, not in the unwanted (global)
one. In particular, if we assume a naive SUSY breaking phase
transition, many horizons choose the unwanted deeper minimum, and the
current horizon contains many regions which dropped into the unwanted
minimum. Notice that it is unclear whether the thermal inflation could
solve this problem, even though the reheating temperature after the
thermal inflation is relatively low. This is because the current
horizon scale contains many different horizons before the thermal
inflation. Therefore, even if the SUSY breaking phase transition
occurs before the thermal inflation, current horizon still contains
many regions of unwanted minimum.

Keeping these arguments in mind, it is important to develop a
cosmologically consistent scenario based on GMSB. Importantly, one
should remember that above problems are usually based on ``minimal''
assumption, and in particular in direct gauge mediation model, the
above arguments do not take account of a possible effect from the SUSY
breaking field.  In the following sections, we will see what happens
if we include its effects.

\section{Model}
 \label{sec:model}
 \setcounter{equation}{0}

In this section, we first briefly review the class of models we are
interested in. As we mentioned in the introduction, we consider a
cosmology of the direct gauge mediation model in which the potential
of the SUSY breaking field is stabilized by the inverted hierarchy
mechanism. In this section, we discuss the general features of such
models~\cite{PRL79-18,NPB510-12,PLB413-336,hph9804450}. An explicit
example of the model is shown in Appendix~\ref{app:example}.

\subsection{Framework}

The model is based on the symmetry ${\rm G=G_S\times G_B\times
G_{SM}}$.  Here, G${\rm _S}$ is the strong gauge interaction whose
dynamics induces the gaugino condensation. On the other hand, G$_{\rm
B}$ is introduced to stabilize the minimum of the potential of the
SUSY breaking field, and G$_{\rm SM}$ is the standard ${\rm
SU(3)_C\times SU(2)_L\times U(1)_Y}$ gauge croup. In this framework,
we introduce the following class of chiral superfields in order to
induce a desired dynamics to break SUSY: the SUSY breaking field
$\Sigma$ which transforms under G$_{\rm B}$, ``messengers''
$q+\bar{q}$ which transform under G${\rm _B}$ and G$_{\rm SM}$, and
``strongly interacting'' fields $Q+\bar{Q}$ which transform under
G${\rm _S}$ and G${\rm _B}$.  These chiral superfields have a
superpotential of the form
 \begin{eqnarray}
  W = y_Q \Sigma \bar{Q} Q + y_q \Sigma \bar{q} q.
 \end{eqnarray}
 Notice that this superpotential has $R$-symmetry which is
non-anomalous for ${\rm G_S}$; under this $R$-symmetry, $\Sigma$ has
charge $+2$. As we will discuss later, this $R$-symmetry is explicitly
broken by supergravity effect.

We construct a model so that $\Sigma$ has a flat direction which is
parameterized by an invariant made out of $\Sigma$. On this flat
direction, G$_{\rm B}$ is broken by the vacuum expectation value (VEV)
of $\Sigma$. The first requirement to the model is that, on this flat
direction, all but one degrees of freedom in $\Sigma$ are eaten by
Higgs mechanism. We parametrize the remaining degrees of freedom by
$X$.

On this flat direction, chiral superfields coupled to $\Sigma$ also
acquire masses. This effect changes the $\beta$-function of G$_{\rm
S}$ at $\mu\sim y_Q\langle\Sigma\rangle$, since $Q$ and $\bar{Q}$ have
mass of $m_Q\sim y_Q\langle\Sigma\rangle$ and decouple at this scale.
As a result, the strong scale for the theory above $m_Q$, $\Lambda$,
differs from that after $Q$ and $\bar{Q}$ decouples, $\Lambda_{\rm
eff}$. These two quantities are related as $\Lambda_{\rm eff}^{3\mu_{\rm
G_S}}=\Lambda^{3\mu_{\rm G_S}-\mu_Q}m_Q^{\mu_Q}$, where $\mu_{\rm
G_S}$ and $\mu_Q$ are Dynkin indices for the adjoint and $Q+\bar{Q}$
representations of G$_{\rm S}$, respectively. The second requirement
to the model is that these Dynkin indices satisfy the relation
$\mu_{\rm G_S}=\mu_Q$, so that the linear superpotential is induced by
the gaugino condensation; below the strong scale, effective
superpotential is induced by the gaugino condensation as
 \begin{eqnarray}
  W_{\rm eff} \sim \Lambda_{\rm eff}^3
  \sim (y_QX)^{\mu_Q/\mu_{\rm G_S}} \Lambda^{3-\mu_Q/\mu_{\rm G_S}}.
 \end{eqnarray}
 Therefore, if $\mu_{\rm G_S}=\mu_Q$, superpotential is linear in $X$,
and supersymmetry is broken by the VEV of the $F$-term of the $X$
field; $F_X\sim y_Q\Lambda^2$. (This class of model of SUSY breaking
is originally discussed in Ref.~\cite{IzaYanInt}.)

With the above superpotential, scalar potential is given by
 \begin{eqnarray}
  V = \frac{|\partial_X W_{\rm eff}|^2}{\partial_X^*\partial_X K}
  \simeq \frac{y_Q^2\Lambda^4}{Z_\Sigma(X^*,X)},
 \label{V_global}
 \end{eqnarray}
 where $\partial_X$ represents the derivative with respect to $X$, $K$
is the K\"ahler potential, and $Z_\Sigma$ is the wave function
renormalization of $\Sigma$. At the tree level, $Z_\Sigma =1$, and
hence $V$ does not depend on $X$.  In this case, VEV of $X$ is
undetermined. However, once we include the radiative corrections, the
situation changes.  Since $\Sigma$ interacts through gauge and Yukawa
interactions, non-trivial K\"ahler potential is induced by radiative
correction. $X$ dependence of $Z_\Sigma$ is governed by the
renormalization group equation (RGE).  At the one loop level, RGE for
$Z_\Sigma$ is given by
 \begin{eqnarray}
  \frac{d\ln Z_\Sigma}{d\ln\mu} = 
  \frac{1}{16\pi^2} 
  \left[ C_{\rm B}g_{\rm B}^2(\mu) - C_Qy_Q^2(\mu) - C_qy_q^2(\mu) 
  \right],
 \label{dZ/dt}
 \end{eqnarray}
 where coefficients $C_{\rm B}$, $C_Q$, and $C_q$ are all positive. If
the $\beta$-function for $Z_\Sigma$ vanishes, we have a minimum (or
maximum) of the potential. The important point is that the gauge and
the Yukawa contributions have opposite signs in the $\beta$-function
of $Z_\Sigma$. As we can see in Eqs.~(\ref{V_global}) and
(\ref{dZ/dt}), gauge contribution makes $Z_\Sigma$ larger at higher
energy and drives $|X|$ to a larger value, while Yukawa contribution
affects in the opposite way. As a result, if the gauge piece dominates
the $\beta$-function in the low energy, and also if the effect of the
Yukawa terms wins in the high energy, $X$ has a minimum at $|X|=v$
where the $\beta$-function vanishes:
 \begin{eqnarray}
  C_{\rm B}g_{\rm B}^2(v) = C_Qy_Q^2(v) + C_qy_q^2(v).
 \end{eqnarray}
 
Therefore, in this class of models, potential for $X$ is stabilized by
the scale dependence of the wave function renormalization factor
$Z_\Sigma$. At the minimum, $F$-component of the $X$ field has a
non-vanishing VEV of $O(y_Q\Lambda^2)$, and SUSY is broken. Then,
fermionic component of $X$ becomes the goldstino, and in supergravity,
it is absorbed in the gravitino. In this case, gravitino mass is
related to $F_X$ as
 \begin{eqnarray}
  m_{3/2} = \frac{F_X}{\sqrt{3}M_*}.
 \end{eqnarray}

Once $X$ gets a VEV $v$, $q$ and $\bar{q}$ acquire a SUSY preserving
mass of $\sim y_qv$, as well as SUSY breaking mass squared for the
scalar component of $\sim y_qF_X$. Since $q$ and $\bar{q}$ have
quantum numbers under the standard model gauge group, the SUSY
breaking masses for the SSM superparticles arise by integrating out
these messenger particles. As in the case of well-known ordinary gauge
mediation model, the ratio,
 \begin{eqnarray}
  B_Q = \frac{F_X}{v},
 \end{eqnarray}
 determines the scale of the SUSY breaking masses in the SSM
superparticles which are of $O(\alpha_{\rm SM}/4\pi)B_Q$, with
$\alpha_{\rm SM}$ being the appropriate gauge coupling of the standard
model gauge group. $B_Q$ should be in the range of $10^4{\rm
~GeV}\lesssim B_Q\lesssim 10^5{\rm ~GeV}$; the lower bound is from
experimental lower bounds on the masses of the superparticles, while
the upper bound is from the naturalness point of view.

\begin{figure}
 \centerline{\epsfxsize=0.5\textwidth\epsfbox{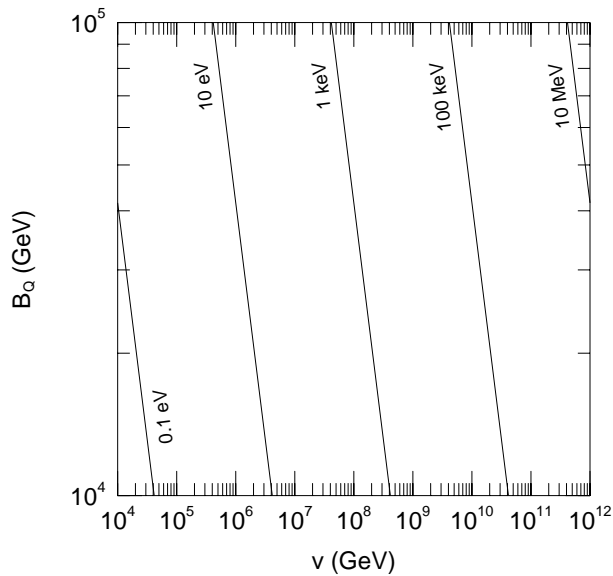}}
 \caption{Contours of the constant gravitino mass on the $v$ vs.~$B_Q$
plane.}
 \label{fig:m_grav}
 \end{figure}

In the following discussions, we use $v$ and $B_Q$ as independent
parameters in the model, and rewrite other quantities as functions of
them. For example, the gravitino mass is given by
 \begin{eqnarray}
  m_{3/2} = \frac{vB_Q}{\sqrt{3}M_*}.  
 \label{m_grav}
 \end{eqnarray}
 In Fig.~\ref{fig:m_grav}, we show the contours of the constant
gravitino mass on the $v$ vs.~$B_Q$ plane.

Before closing this subsection, we discuss the allowed range of $v$.
Recalling that $\sim y_q^2v^2$ and $\sim y_qF_X$ are the diagonal and
off-diagonal elements of the mass squared matrix of the messenger
scalars, $v$ has to be larger than $\sim y_q^{-1/2}F_X^{1/2}$ for the
positivity of the eigenvalues of the mass squared matrix; otherwise,
messenger scalar has a VEV and the standard model gauge groups are
broken.  Another constraint is from the stability of the SUSY breaking
minimum.  In some class of models, there is a SUSY preserving true
vacuum at the origin ($|X|=0$). In this case, the tunneling rate from
the SUSY breaking vacuum to the true vacuum has to be small enough, so
that the lifetime of the SUSY breaking vacuum is longer than the age
of the Universe. This issue is discussed in Ref.~\cite{NPB510-12}, and
it requires $v/\Lambda\gtrsim 10$. Since $F_X\sim y_Q\Lambda^2$,
$v\gtrsim 10y_Q^{-1/2}F_X^{1/2}$ is required in models with true
vacuum at the origin. Notice that this constraint is more stringent
than the one from the stability of the messenger potential, if
$y_Q\sim y_q$.

One important lower bound on $v$ is derived for the validity of our
perturbative approach. If $v$ is close to the strong scale $\Lambda$,
${\rm SU(2)_S}$ dynamics generates non-perturbative K\"ahler
potential, and our perturbative arguments break down.  At the scale
$\Lambda_{\rm eff}$, we expect a non-perturbative contribution to the
K\"ahler potential of the form $\delta K\sim [({\cal W^\alpha
W_\alpha})(\bar{\cal W}_{\dot{\alpha}}\bar{\cal
W}^{\dot{\alpha}})]^{1/3}\sim \Lambda_{\rm eff}^*\Lambda_{\rm
eff}$~\cite{PLB113-231,NPB510-12}, where ${\cal W}^\alpha$ is the
spinor superfield for ${\rm SU(2)_S}$. With the naive dimensional
analysis~\cite{NDA}, coefficient of this operator is estimated to be
of $O((4\pi)^{-2/3})$. Requiring this effect to be smaller than the
perturbative one, $v$ has to be larger than
$10^9-10^{10}$~GeV~\cite{NPB510-12}. However, it may be possible that
the minimum of the potential exists even with the non-perturbative
effect. (For a model in which the SUSY breaking minimum is stabilized
by non-perturbative effects, see Ref.~\cite{nonpert}.) In this case,
we need to make some dynamical assumptions not to change our following
discussions, since the non-perturbative effect is not calculable.
However, most importantly, strong dynamics does not break the
$R$-symmetry, and hence the properties of the $R$-axion is unchanged.
Consequently, evolution of the $R$-axion is basically unaffected, even
if the non-perturbative effect becomes important at the minimum of the
potential. With these caveats in mind, we also consider the case with
$v\lesssim 10^9{\rm ~GeV}$ in the following discussion. In this case,
we make an implicit assumption that the minimum of the potential is
stabilized even after taking account of the non-perturbative effect.

Finally, we comment on the upper bound on $v$. As can be seen from
Eq.~(\ref{m_grav}), gravitino mass becomes larger than $\sim 100{\rm
~GeV}$ for $v\gtrsim 10^{16}{\rm ~GeV}$ with $B_Q\sim 10^4{\rm ~GeV}$.
In this case, supergravity contribution to the sfermion masses becomes
comparable to the gauge mediation piece, and hence SUSY FCNC problem
arises again. Therefore, $v\ll 10^{16}{\rm ~GeV}$ is required to solve
SUSY FCNC problem.

\subsection{Potential of $X$}

In order to discuss the cosmological implication of $X$, it is
important to understand the properties of its potential, which is the
subject of this subsection.

We are interested in the models in which the minimum of the $X$ field
is determined by the inverted hierarchy mechanism. In this class of
models, the potential at the global level is only lifted by the
renormalization group effects, and hence it can be expanded by powers
of $\ln X^*X$:\footnote
 {After the second line, we omit the constant piece. We assume that
the cosmological constant is cancelled out by supergravity effect, and
hence the constant term does not matter.}
 \begin{eqnarray}
  V_{\rm global} &=& F_X^2 / Z_\Sigma
 \nonumber \\ &=&
  F_X^2 \left[ 
  -\frac{1}{16\pi^2} \frac{Z^{\prime}_\Sigma (\mu)}{Z_\Sigma^2(\mu)} 
  \ln\frac{X^*X}{\mu^2}
  +\frac{1}{(16\pi^2)^2}
  \left( \frac{Z_\Sigma^{\prime 2}(\mu)}{Z_\Sigma^3(\mu)} 
  - \frac{Z_\Sigma^{\prime\prime}(\mu)}{Z_\Sigma^2(\mu)} \right)
  \left(\ln\frac{X^*X}{\mu^2}\right)^2
  + \cdots \right]
 \nonumber \\ &\equiv&
  F_X^2 \left[
  \frac{\zeta_1 (\mu)}{16\pi^2} \ln\frac{X^*X}{\mu^2}
  + \frac{\zeta_2 (\mu)}{(16\pi^2)^2}
  \left(\ln\frac{X^*X}{\mu^2}\right)^2
  + \cdots \right],
 \label{V_expansion}
 \end{eqnarray}
 with
 \begin{eqnarray}
  Z^{\prime}_\Sigma \equiv 8\pi^2 \frac{dZ_\Sigma}{d\ln\mu},~~~
  Z^{\prime\prime}_\Sigma \equiv 
  (8\pi^2)^2 \frac{d^2Z_\Sigma}{d(\ln\mu)^2}.
 \end{eqnarray}
 Here, $\zeta_n$ is a $2n$-th polynomial of gauge and Yukawa coupling
constants. We factorized relevant powers of $16\pi^2$, so that
coefficients in $\zeta_n$ become typically of $O(1)$. Without
cancellation, $\zeta_n$ is close to 1 if some of the gauge or Yukawa
coupling is close to 1.

In Eq.~(\ref{V_expansion}), $\mu$ can be an arbitrary scale; $\mu$
dependence is cancelled out by the renormalization group effect. If we
take $\mu =v$ where $Z^{\prime}_\Sigma$ vanishes, $V_{\rm global}$
starts with a term which is quadratic in $\ln X^*X$, and hence the
potential has an extremum at $|X|=v$. In particular, if
$\zeta_2(v)>0$, potential has a minimum there, which is what we
desired. Hereafter, we assume that the gauge and Yukawa coupling
constants are arranged so that $\zeta_2(v)>0$.  Around this minimum,
potential starts with $(\ln X^*X)^2$ term which is suppressed by
$(16\pi^2)^{-2}$.  However, once $|X|$ becomes much larger (or
smaller) than $v$, $\zeta_1$ at that scale may become large so that
the potential is approximated to be linear in $\ln X^*X$, with being
suppressed only by $(16\pi^2)^{-1}$.

So far, we have discussed the potential in the framework of global
SUSY. However, supergravity effect also generates important terms in
the potential. First of all, all the scalar fields receive SUSY
breaking masses of the order of the gravitino mass. This effect
becomes important especially when the amplitude of $X$ becomes large.
Another important implication is that the $R$-symmetry is (explicitly)
broken due to the supergravity effect if the cosmological constant is
cancelled out~\cite{NPB426-3}. Under U(1)$_R$ symmetry, superpotential
has the charge of $+2$. However, in order to cancel the cosmological
constant, a constant term is needed in the superpotential.  From the
cross term between them, $R$-symmetry breaking potential is induced:
 \begin{eqnarray}
  V_{\not{R}} \sim - \frac{F_X^2}{M_*}
  (X^*+X) \times f(X^*X/M_*^2),
 \label{V_Rbrk}
 \end{eqnarray}
 where $f$ is an unknown function. We expand this function as
 \begin{eqnarray}
  f(x)=k_0+k_1x+\cdots,
 \label{f_expand}
 \end{eqnarray}
 where coefficients $k_n$ are expected to be of $O(1)$. By combining
these contributions, supergravity contribution is of the form:
 \begin{eqnarray}
  V_{\rm SUGRA} \sim m_{3/2}^2 X^*X + V_{\not{R}}.
 \end{eqnarray}

In fact, the linear term in $V_{\not{R}}$ may cause a problem.  If we
neglect the logarithmic terms, non-vanishing $k_0$ shifts the minimum
of the potential from $X=0$ to $X\sim k_0M_*$. Thanks to the
logarithmic term in $V_{\rm global}$, potential can have a minimum at
$|X|=v\ll M_*$.  However, with the potential
 \begin{eqnarray}
  V \sim \frac{\zeta_1}{16\pi^2}F_X^2 \ln X^*X
  - k_0 \frac{F_X^2}{M_*} (X+X^*)
  + m_{3/2}^2 X^*X + \cdots,
 \end{eqnarray}
 another minimum still exists at $|X|\sim M_*$ when $k_0\sim O(1)$.
With this minimum, $X$ is more likely to settle down to this unwanted
minimum if $X$ has an initial amplitude of $O(M_*)$.  This is because
the potential is dominated by the supergravity contribution for such a
large amplitude, and hence $X$ does not feel the effect of the
logarithmic piece unless $|X|$ becomes small enough.  In order to
remove this unwanted minimum, we assume $k_0$ to be small enough,
 \begin{eqnarray}
  k_0\lesssim \frac{\zeta_1}{4\pi}.
 \end{eqnarray}
 If $\zeta_1$ is of $O(1)$, this is a tuning of 10~\% level, and we
believe accidental cancellation would be enough for this
suppression.\footnote
 {The values of $k_0$ at $|X|\sim M_*$ and at $|X|\sim v$ are supposed
to be different. In particular, in our case, moduli field exists which
has a very large initial amplitude of $O(M_*)$. The value of $k_0$
should be affected by the evolution of the moduli field.}

\subsection{Properties of the Physical Modes}

At around the minimum ($|X|\sim v$), we have two physical scalars from
$X$. In order to discuss the properties of these fields, it is
convenient to parametrize the $X$ field as
 \begin{eqnarray}
  X = \left(v + \frac{1}{\sqrt{2}}\sigma \right) e^{ia/\sqrt{2}v}.
 \end{eqnarray}
 Expanding the potential around the minimum, we obtain the mass of the
$\sigma$ as
 \begin{eqnarray}
  m_\sigma^2 = - \frac{Z_\Sigma^{\prime\prime}(v)}{(16\pi^2)^2} 
  \frac{F_X^2}{v^2}
  = \frac{\zeta_2(v)}{(16\pi^2)^2} \frac{F_X^2}{v^2},
 \label{m_s}
 \end{eqnarray}
 where we normalized as $Z_\Sigma(v)=1$. Notice that $\sigma$ is as
heavy as the SSM superpartners if all the gauge and Yukawa coupling
constants are of the same order. On the contrary, $a$ is the
pseudo-Nambu-Goldstone boson for the $R$-symmetry, which is usually
called $R$-axion.  The main source of the $R$-axion mass is the
$R$-symmetry breaking term from the supergravity effect;
 \begin{eqnarray}
  V_{\not{R}} = - k_0 \frac{F_X^2}{M_*} (X^*+X),
 \label{V_Rbrk1}
 \end{eqnarray}
 where $k_0$ is (unknown) $O(1)$ constant introduced in
Eqs.~(\ref{V_Rbrk}) and (\ref{f_expand}). (Around the minimum $|X|=v$,
higher order terms are suppressed by powers of $v^2/M_*^2$.) From this
potential, the $R$-axion mass is given by~\cite{NPB426-3}\footnote
 {There is another contribution to the $R$-axion mass from the QCD
anomaly, which is, however, much smaller than the supergravity effect.
This fact suggests that it is difficult to use this $R$-axion as a
solution to the strong CP problem, unless the supergravity effect is
much smaller than the naive expectation.}
 \begin{eqnarray}
  m_a^2 = k_0 \frac{F_X^2}{vM_*}
  \simeq 
  \left[ 6~{\rm GeV} \times k_0^{1/2}
  \left( \frac{B_Q}{\rm 10^5~GeV} \right)
  \left( \frac{v}{\rm 10^{10}~GeV} \right)^{1/2}
  \right]^2.
 \label{ma}
 \end{eqnarray}

Next, we consider the decay rate of these fields. The $R$-axion
dominantly decays into gauge boson pairs.\footnote
 {If the ordinary quarks and leptons have non-vanishing $R$-charge,
the $R$-axion may decay into these fermions. However, it is unknown
whether the $R$-symmetry can be consistently defined in the SSM
sector, and the couplings of the $R$-axion to the quarks and leptons
are model-dependent. In particular, these couplings cannot be fixed
unless we specify the mechanism to generate so-called $\mu$- and
$B$-parameters. Furthermore, these decay processes are chirality
suppressed. In this paper, we assume that these decay modes are
negligibly small. The $R$-axion may also decay into gaugino
pairs. However, the $R$-axion is lighter than the gauginos in most of
the parameter region we consider.  Therefore, we do not consider this
decay mode. Notice that the decay rate for this process is at most
comparable to that into gauge bosons.  Therefore, qualitatively, the
following arguments are unchanged even if the $R$-axion decays into
gaugino pairs.}
 Since the $R$-axion couples to the messenger multiplets $q$ and
$\bar{q}$ which transform under ${\rm SU(3)_C\times SU(2)_L\times
U(1)_Y}$, the $R$-axion has a coupling to the standard model gauge
bosons at the one-loop level. Then, its decay rate is calculated as
 \begin{eqnarray}
  \Gamma_a = \frac{n_{\rm G}}{32\pi}
  \left(\frac{\alpha_{\rm SM}b_q}{4\pi}\right)^2
  \frac{m_a^3}{v^2}.
 \label{Gamma_a}
 \end{eqnarray}
 Here, $n_{\rm G}$ is the number of the final state, $b_q$ is the
$\beta$-function coefficient of $q+\bar{q}$, and $\alpha_{\rm SM}$ is
the corresponding gauge coupling constant.  For example, for the decay
into the gluon pair ($a\rightarrow gg$), $n_{\rm G}=8$, $b_q=N_5$, and
$\alpha_{\rm SM}=\alpha_{\rm s}$, while for the decay into the photon
pair ($a\rightarrow\gamma\gamma$), $n_{\rm G}=1$,
$b_q=\frac{8}{3}N_5$, and $\alpha_{\rm SM}=\alpha_{\rm em}$, with
$N_5$ being the number of messenger chiral superfields in units of
${\bf 5}+{\bf \bar{5}}$ representations of SU(5). (For the model shown
in Appendix~\ref{app:example}, $N_5=2$.)

On the other hand, the $\sigma$ field dominantly decays into the
$R$-axion pair ($\sigma\rightarrow aa$). From the following
Lagrangian:
 \begin{eqnarray}
  {\cal L} =
  \partial_\mu X^* \partial_\mu X \simeq
  \frac{1}{2} \partial_\mu \sigma \partial_\mu \sigma
  + \frac{1}{2} \partial_\mu a \partial_\mu a 
  + \frac{1}{\sqrt{2}v} \sigma \partial_\mu a \partial_\mu a 
  + \cdots,
 \end{eqnarray}
 we obtain the decay rate of this process to be
 \begin{eqnarray}
  \Gamma_\sigma = \frac{1}{64\pi} \frac{m_\sigma^3}{v^2}.
 \label{Gamma_s}
 \end{eqnarray}
 Notice that $\sigma$ also decays into the gauge boson pairs, and the
decay rate for this process is given by a similar formula as
Eq.~(\ref{Gamma_a}) with $m_a$ being replaced by $m_\sigma$. Comparing
the decay rates for these processes, we can see that
$\sigma\rightarrow aa$ is the dominant decay mode for $\sigma$.

The decay mode into the gravitino pair is potentially significant.
Indeed, the SUSY breaking field has an interaction as
 \begin{eqnarray}
  {\cal L} =  
  - \frac{1}{2M_*^2} W \bar{\psi}_\mu \sigma^{\mu\nu} P_L \psi_\nu
  + {\rm h.c.} \simeq
  - \frac{1}{2M_*^2} F_X X \bar{\psi}_\mu \sigma^{\mu\nu} P_L \psi_\nu
  + {\rm h.c.},
 \end{eqnarray}
 with $\psi_\mu$ being gravitino. This interaction induces the decay
process $X\rightarrow\psi_\mu\psi_\mu$, and in particular, the decay
into the longitudinal component is the most important. With the
Lagrangian given above, we obtain the decay rate as
 \begin{eqnarray}
  \delta\Gamma_X \equiv \Gamma (X\rightarrow\psi_\mu\psi_\mu)
  = \frac{1}{32\pi} \frac{m_X^5}{F_X^2},
 \label{Gamma_X(grav)}
 \end{eqnarray}
 with $X=a$ and $\sigma$. Comparing Eq.~(\ref{Gamma_X(grav)}) with
Eqs.~(\ref{Gamma_a}) and (\ref{Gamma_s}), the branching ratio for the
process $X\rightarrow\psi_\mu\psi_\mu$ is estimated as
 \begin{eqnarray}
  && {\rm Br} (a\rightarrow\psi_\mu\psi_\mu) \simeq 
  \frac{\delta\Gamma_a}{\Gamma_a}\simeq
  \frac{1}{n_{\rm G}}
  \left( \frac{m_a}{(\alpha_{\rm SM}b_q/4\pi)B_Q}\right)^2,
 \label{Br_a} \\
  && {\rm Br} (\sigma\rightarrow\psi_\mu\psi_\mu) \simeq
  \frac{\delta\Gamma_\sigma}{\Gamma_\sigma}\simeq
  2 \left( \frac{m_\sigma}{B_Q}\right)^2.
 \label{Br_s}
 \end{eqnarray}
 Notice that these branching ratios are much smaller than 1. In the
parameter region we consider, the $R$-axion mass is usually smaller
than $O(\alpha_{\rm SM}/4\pi)B_Q$.  Furthermore, as shown in
Eq.~(\ref{m_s}), $m_\sigma$ is smaller than $B_Q$.  Therefore, in both
cases, the decay modes into the gravitino pair are suppressed. Even
with small branching ratios, however, the decay of the $R$-axion and
$\sigma$ may overproduce the gravitino, resulting in too much mass
density of the Universe. We will come back to this point in the next
section.

\section{Evolution of the Scalar Fields}
 \label{sec:evolution}
 \setcounter{equation}{0}

Now, we are ready to discuss the evolutions of the scalar fields,
i.e., the SUSY breaking field $X$ and the moduli field $\phi$. Since
the behavior of $X$ changes at $|X|\sim v$, we first consider the
evolution when $|X|\gg v$. Then, the behavior of $X$ when $|X|\sim v$
is considered.

\subsection{$|X|\sim M_*$}

In our discussion, we adopt the picture of the inflationary cosmology;
we assume a primordial inflation which solves horizon and flatness
problems. During this inflation, $X$ and $\phi$ are shifted from their
minimum, and these fields have very large initial amplitudes. We
assume that their initial amplitudes are both of $O(M_*)$.  This may
happen because of the chaotic assumption on the initial condition of
scalar fields~\cite{PLB129-177}, due to modification of the scalar
potential during the inflation with large expansion
rate~\cite{DinRanTho}, or in the case of no-scale type
supergravity~\cite{hph9805300}. Importantly, $e$-folding of this
primordial inflation is large enough to solve the horizon problem, and
hence $X$ and $\phi$ have coherent initial states for the scale of the
current horizon. After the primordial inflation, reheating happens,
and the Universe becomes radiation dominated. At this stage, $X$ and
$\phi$ keep their initial values as far as the expansion rate is
larger than their masses.

Once the expansion rate of the Universe becomes comparable to the mass
of $X$ and $\phi$, these fields start to move. Since these fields (in
particular, $X$) are assumed to have large amplitudes of $O(M_*)$, the
potential for these fields are initially dominated by the supergravity
contributions. Furthermore, with these large initial amplitudes,
energy density of the radiation becomes comparable to those of $X$ and
$\phi$ when these fields start to move. Then, after this stage,
energy density of radiation decreases faster than those of $X$ and
$\phi$, and the Universe is dominated by the coherent oscillation.

Once the moduli field $\phi$ starts oscillation, its evolution is
quite simple. Approximating its potential to be quadratic as $\sim
m_{3/2}^2 \phi^2$, the energy density of the moduli field scales as
$R^{-3}$, with $R$ being the scale factor. 

One may worry about a possible shift of the minimum of the moduli
potential, in particular, due to the non-vanishing expansion rate $H$
induced by $X$. When the Universe is dominated by a condensation of a
scalar field (in our case, $X$), supergravity effect induces extra
terms in the moduli potential, which are proportional to
$H^2$~\cite{DinRanTho}. This effect shifts the minimum of the moduli
potential, and $\phi$ oscillates around the shifted minimum.  If the
minimum moves more drastically than $\phi$, behavior of $\phi$ is
affected by the change of the potential, and our argument breaks
down. However, in our case, this is not the case; time scale of the
shift of the minimum is $H^{-1}$, which is much longer than that of
the oscillation $m_{3/2}^{-1}$. As a result, $\phi$ can catch up with
the shift of the minimum. Therefore, if we consider the oscillation
around the shifted minimum, our argument is unchanged. In particular,
this does not affect the calculation of the current energy density of
the moduli field since the shifted minimum smoothly approaches to the
true minimum as the Universe expands. (For details, see
Appendix~\ref{app:shift}.)

Evolution of the SUSY breaking field $X$ is more complicated. When $X$
starts to move, angular component of $X$ ($R$-axion mode) is excited
by the Affleck-Dine mechanism and $R$-number is generated.
Furthermore, when logarithmic potential takes over the supergravity
contribution, energy density of $X$ decreases much slower than that of
the moduli field.

Let us first discuss the generation of the $R$-number due to the
Affleck-Dine mechanism~\cite{NPB249-361}. For our argument, it is
convenient to define the $R$-number density:
 \begin{eqnarray}
  n_R = i(X^*\dot{X} - \dot{X}^*X).
 \end{eqnarray}
 Then, with the $R$-symmetry breaking terms given in
Eq.~(\ref{V_Rbrk}), time evolution of $n_R$ is given by
 \begin{eqnarray}
  \dot{n}_R + 3Hn_R &=& 
  - i \left( \frac{\partial V}{\partial X}X - 
  \frac{\partial V}{\partial X^*}X^* \right)
 \nonumber \\ &=&
  2F_X^2{\rm Im} (X/M_*) \times f(|X|^2/M_*^2).
 \label{dotn_r}
 \end{eqnarray}
 Notice that, if the potential respects the $R$-symmetry, the
right-hand side of Eq.~(\ref{dotn_r}) vanishes.

In our case, $R$-symmetry breaking effect is the largest when $X$ has
the maximum amplitude, and $R$-number asymmetry is generated when $X$
starts to move (see Appendix~\ref{app:nR}). $R$-number at this stage
is estimated as
 \begin{eqnarray}
  n_R \sim H^{-1}\times 2F_X^2{\rm Im} (X_0/M_*) \times f(|X_0|^2/M_*^2).
 \end{eqnarray}
 By using $H\sim m_{3/2}$, we obtain
 \begin{eqnarray}
  n_R \sim \frac{2 F_X^2\xi \sin\theta_0}{m_{3/2}},
 \end{eqnarray}
 where the initial value of $X$ is parameterized as
$X_0=|X_0|e^{i\theta_0}$, and $\xi\sim (|X_0|/M_*)\times
f(|X_0|^2/M_*^2)$ is expected to be of $O(1)$ if $|X_0|\sim M_*$.

\begin{figure}
 \centerline{\epsfxsize=0.5\textwidth\epsfbox{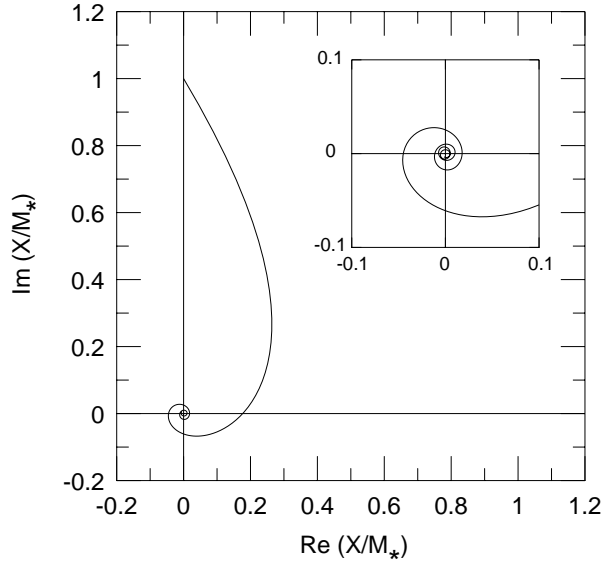}}
 \caption{Initial motion of the SUSY breaking field $X$ on the complex
$X$ plane. We choose the set of parameters as $X_0=M_*e^{i\pi/2}$,
$\zeta_1 = 0.3$, $k_1 = 0.2$, and other $\zeta_n$'s and $k_n$'s are
taken to be 0.}
 \label{fig:motion}
 \end{figure}

In Fig.~\ref{fig:motion}, we show a typical behavior of $X$ on the
complex $X$ plane. As an example, we choose the set of model
parameters as
 \begin{eqnarray}
  X_0 = M_*e^{i\pi/2}, ~~~\zeta_1 = 0.3, ~~~k_1 = 0.2, 
 \end{eqnarray}
 and other $\zeta_n$'s and $k_n$'s are taken to be 0.

In order to discuss the evolutions of the moduli field and $X$
simultaneously, it is convenient to take the ratio of the energy
density of $\phi$, $\rho_\phi$, to $n_R$. Since the initial energy
density of the moduli field is of the order of $m_{3/2}^2\phi_0^2\sim
F_X^2\times (\phi_0/M_*)^2$, we obtain
 \begin{eqnarray}
  \frac{\rho_\phi}{n_R} 
  \sim \frac{m_{3/2}}{2\xi\sin\theta_0} 
  \left(\frac{\phi_0}{M_*}\right)^2.
 \label{phophi/nr}
 \end{eqnarray}
 Notice that $\rho_\phi$ and $n_R$ are both proportional to $R^{-3}$,
and the ratio $\rho_\phi /n_R$ remains constant as far as the
$R$-symmetry breaking effects can be neglected.

\subsection{$v\lesssim |X|\lesssim M_*$}

Once the SUSY breaking field $X$ and the moduli field $\phi$ start to
oscillate, their amplitudes adiabatically decrease with the expansion
of the Universe. Their relation is determined by
Eq.~(\ref{phophi/nr}), and the final result can be derived without
discussing the detail of their evolutions. However, in this
subsection, we discuss how their amplitudes behave when $v\lesssim
|X|\lesssim M_*$ for a better understanding of the behaviors of $X$
and $\phi$.

As discussed in Appendix~\ref{app:osc}, amplitude of the scalar field
$\varphi$ (corresponding to $X$ and $\phi$) obeys the relation
 \begin{eqnarray}
  m_{\rm eff} \varphi^2 R^3 = {\rm const.},
 \end{eqnarray}
 where the ``effective mass'' $m_{\rm eff}$ is defined as
 \begin{eqnarray}
  m_{\rm eff}^2 = \frac{1}{2}
  \left( \frac{1}{\varphi^*} \frac{\partial V}{\partial\varphi} + 
  \frac{1}{\varphi} \frac{\partial V}{\partial\varphi^*} \right).
 \end{eqnarray}
 
Effective masses of $X$ and $\phi$ depend on their amplitudes
differently. Potential for the moduli field $\phi$ is parabolic, and
hence $m_{\rm eff}$ is independent of the amplitude.  Thus, the
amplitude of the moduli field scales as $R^{-3/2}$.

We next discuss the evolution of $X$. For this purpose, let us remind
the structure of the potential of $X$:\footnote
 {We neglect the $R$-symmetry breaking part; even if we include its
effect, the result is unchanged.}
 \begin{eqnarray}
  V \sim \frac{\zeta_1}{16\pi^2} F_X^2 \ln X^*X
  + m_{3/2}^2 X^*X.
 \end{eqnarray}
 As mentioned in the previous subsection, supergravity contribution
wins the global SUSY contribution when $|X|$ is large. Comparing two
contributions, supergravity effect is more important above a threshold
value $X_{\rm thr}$ which is estimated as
 \begin{eqnarray}
  X_{\rm thr} \sim \frac{\sqrt{\zeta_1}}{4\pi} M_*.
 \end{eqnarray}
 For $|X|\gtrsim X_{\rm thr}$, supergravity effect wins the global
SUSY contribution, and $m_{\rm eff}\sim m_{3/2}$ for $X$.  In this
case, the amplitude of $X$ scales as $R^{-3/2}$.  On the other hand,
for $|X|\lesssim X_{\rm thr}$, potential is dominated by the
logarithmic piece, and $m_{\rm eff}\propto |X|^{-1}$. In this case,
the amplitude of $X$ is proportional to $R^{-3}$.\footnote
 {If $|X|$ becomes close to $v$, potential is approximately
proportional to $(\ln X^*X)^2$. In this case, this relation receives a
logarithmic correction; $|X|(\ln |X|)^{1/2}R^3={\rm const.}$}

Comparing the evolutions of $X$ and $\phi$, their amplitudes are
related as
 \begin{eqnarray}
  \phi / |X| \sim 
  \left\{
  \begin{array}{ll}
  {\phi_0 / |X_0|} & {:~~~|X|\gtrsim X_{\rm thr}} \\
  {(X_{\rm thr}/|X|)^{1/2}\times (\phi_0/|X_0|)}
  & {:~~~|X|\lesssim X_{\rm thr}}
  \end{array}\right. ,
 \label{rel_XvsPhi}
 \end{eqnarray}
 with $\phi_0$ and $X_0$ being the initial values of $X$ and $\phi$,
respectively.  For $|X|\gtrsim X_{\rm thr}$, $X$ and $\phi$ scales in
the same way, while for $|X|\lesssim X_{\rm thr}$, the amplitude of
$X$ decreases faster than that of $\phi$.

The evolution of the energy density of $X$ and $\phi$ is also
non-trivial. When $|X|\gtrsim X_{\rm thr}$, both $X$ and $\phi$ obey
parabolic potential, and their energy density scale as $R^{-3}$.
However, once the amplitude of $X$ becomes smaller than $X_{\rm thr}$,
potential for $X$ is lifted only logarithmically, and energy density
of $X$ decreases very slowly with the decrease of $|X|$. The important
point is that, once the potential of $X$ is dominated by the
logarithmic piece, energy density of $\phi$ decreases much faster than
that of $X$, and hence the energy density of the Universe is dominated
by $X$. Therefore, when $X$ decays, there can be a large entropy
production to dilute the moduli field away.

When the amplitude of $X$ becomes of $O(v)$, $X$ is trapped in the
minimum of the potential. Then, the amplitude of $X$ is fixed to be
$v$, and we need to consider the excitations around the minimum. They
are phase degrees of freedom (i.e., $R$-axion) and radial degrees of
freedom. Effects of these fields are discussed in the following
subsections.

Some of the models have a SUSY preserving true vacuum at the origin
$X=0$, and one may worry whether $X$ can be smoothly trapped in the
SUSY breaking (false) vacuum of $|X|=v$.  If $X$ would overshoot down
to the origin, this scenario would not be phenomenologically viable.
Since $X$ follows an elliptic trajectory on the complex $X$ plane, as
shown in Fig.~\ref{fig:motion}, $X$ can result in the SUSY breaking
minimum at $|X|=v$ without being affected by the potential for $|X|\ll
v$. Important point is that, thanks to the $R$-number conservation in
the comoving volume, the orbit of $X$ is always elliptic once the
$R$-number is generated. Consequently, even though the amplitude of
$X$ decreases with the expansion of the Universe, orbit does not pass
by the origin if enough $R$-number asymmetry is generated. In this
case, when $|X|\sim v$, $X$ traces a trajectory which is
(approximately) parallel to the minimum of the potential ($|X|=v$),
and $X$ is smoothly trapped in the SUSY breaking vacuum. Therefore,
once enough $R$-number asymmetry is generated, $X$ is expected to
result in the minimum of the potential $|X|=v$, irrespective of the
structure of the potential for $|X|\ll v$. Of course, this scenario
depends on the initial value of the $R$-number asymmetry. Importantly,
if the initial amplitude of $X$ is as large as $M_*$, Affleck-Dine
mechanism can generate very large asymmetry, as discussed in the
previous subsection. With a reasonable choice of the model parameters,
we can easily have an elliptic trajectory of $X$ with eccentricity
close to 0. This fact helps us to understand how $X$ can be trapped in
the SUSY breaking vacuum at $|X|=v$.

\subsection{Effect of the $R$-axion}

Once the amplitude of $X$ becomes of $O(v)$, the $X$ field is trapped
in the minimum $|X|=v$. After this stage, the imaginary part
($R$-axion) and the real part of $X$ behave differently. Thus, we
discuss their effects separately.

First, we argue the effect of the $R$-axion $a$. Since $a$ is the
pseudo-Nambu-Goldstone boson, its potential is approximately flat with
slight perturbation due to $V_{\not{R}}$. When we can neglect the
$R$-symmetry breaking effects, motion of $a$ corresponds to the phase
rotation of $X$. At this period, it is convenient to parametrize the
$X$ field as
 \begin{eqnarray}
  X = v e^{i\omega t}.
 \end{eqnarray}
 Here, we neglect the real part of the $X$ field. Then, the $R$-number
is given by
 \begin{eqnarray}
  n_R = 2\omega v^2 = 2\rho_a / \omega,
 \label{nR=2omgv2}
 \end{eqnarray}
 where $\rho_a$ is the energy density of the $R$-axion. With the
expansion of the Universe, $\omega$ decreases adiabatically; since the
$R$-number in the comoving volume is conserved, $\omega$ scales as
$R^{-3}$.\footnote
 {For a scalar field $\varphi$ with flat potential, equation of motion
is given by $\ddot{\varphi}+3H\dot{\varphi}=0$. By solving this
equation, we obtain $\dot{\varphi}R^3={\rm const.}$ Evolution of
$\omega$ is consistent with this relation, if we re-interpret
$e^{i\omega t}\rightarrow e^{ia/\sqrt{2}v}$.}

Once the energy density of the $R$-axion becomes less than the
difference of the potential energy $\Delta V$ on the circle $|X|=v$,
the $R$-axion starts to oscillate around its minimum. Once this
happens, $R$-symmetry is not a good symmetry any more, and the
$R$-axion approximately obeys the parabolic potential. After this
stage, energy density of $a$ scales as $R^{-3}$. With the potential
given in Eq.~(\ref{V_Rbrk1}), the difference of the potential energy
is given by $\Delta V=V(X=-v)-V(X=v)=4m_a^2v^2$, and hence two cases
should be matched when $\omega\sim O(m_a)$. Connecting two cases at
$\omega=\omega_{\rm c}$, we obtain
 \begin{eqnarray}
  \frac{\rho_\phi}{\rho_a} \sim 
  \frac{\rho_\phi}{(\omega_{\rm c}n_R/2)}
  \sim
  \frac{m_{3/2}}{\omega_{\rm c}\xi\sin\theta_0}
  \left(\frac{\phi_0}{M_*}\right)^2.
 \label{ratiorho_a}
 \end{eqnarray}
 This ratio remains constant until the $R$-axion decays.\footnote {If
the amplitude of $X$ is largely fluctuated, domain wall is produced
when the $R$-axion gets trapped in the minimum of the potential. Such
a fluctuation is generated during the primordial inflation, and the
domain wall production may be effective if the expansion rate during
the inflation is larger than of $O(v)$~\cite{PRD56-7597}. Therefore,
if we adopt a primordial inflation with relatively small expansion
rate, the domain wall production can be evaded. Furthermore, even if
the domain wall production is effective, collapse of the domain wall
results in semi-relativistic $R$-axions with averaged energy of $\sim
3m_a$~\cite{PRD50-4821}. Therefore, domain wall production does not
modify the ratio $\rho_\phi/\rho_a$ given in Eq.~(\ref{ratiorho_a}) so
much, and the dilution factor calculated below is almost unchanged.}

Motion of the $R$-axion at $\rho_a\sim m_a^2v^2$ is complicated, and
analytic estimation of $\omega_{\rm c}$ is difficult. In our analysis,
we numerically followed the evolution of the $R$-axion, and estimated
the value of $\omega$ for matching. As a result of the numerical
calculation, we found that two cases should be connected with
 \begin{eqnarray}
  \omega = \omega_{\rm c} \simeq 4 m_a.
 \end{eqnarray}
 In the following discussion, we use $\omega_{\rm c}=4m_a$.

When the expansion rate of the Universe becomes comparable to the
decay rate of the $R$-axion, the $R$-axion decays and the Universe is
reheated. By using the instantaneous decay approximation, the
reheating temperature $T_{\rm R}$ is estimated as
 \begin{eqnarray}
  T_{\rm R} &\sim& \left(\frac{\pi^2g_*}{90}\right)^{-1/4}
  \sqrt{\Gamma_a M_*}
 \nonumber \\ &\sim&
  20 {\rm ~MeV}\times
  k_0^{3/4}
  \left( \alpha_{\rm SM}b_qn_{\rm G}^{1/2} \right)
  \left( \frac{g_*}{10} \right)^{-1/4}
  \left( \frac{B_Q}{\rm 10^5~GeV} \right)^{3/2}
  \left( \frac{v}{\rm 10^{10}~GeV} \right)^{-1/4},
 \end{eqnarray}
 where we used the decay rate given in Eq.~(\ref{Gamma_a}) in the
second line. (A special case where this may be irrelevant will be
discussed later.) At this stage, the energy density of the $R$-axion
is converted to that of the radiation, and large amount of entropy is
produced. At the decay time, energy density of radiation is given by
$\rho_{\rm rad}=\rho_a$, and hence
 \begin{eqnarray}
  \frac{\rho_\phi}{s} \sim \frac{3}{4}T_{\rm R}\times 
  \frac{m_{3/2}}{4m_a\xi\sin\theta_0}
  \left(\frac{\phi_0}{M_*}\right)^2,
 \label{rho2s_a}
 \end{eqnarray}
 where $s$ is the entropy density which is related to the energy
density of the radiation as $\rho_{\rm rad}=\frac{3}{4}Ts$. We compare
Eq.~(\ref{rho2s_a}) with the ratio of the critical density to the
entropy density in the present Universe:
 \begin{eqnarray}
  \frac{\rho_{\rm c}}{s} \simeq
  3.6 \times 10^{-9} {\rm ~GeV} \times h^2,
 \label{rhoc2s}
 \end{eqnarray}
 and we obtain
 \begin{eqnarray}
  \Omega_\phi h^2 \sim 40 \times
  k_0^{1/4}
  \left( \alpha_{\rm SM}b_qn_{\rm G}^{1/2} \right)
  \left( \frac{g_*}{10} \right)^{-1/4}
  \left( \frac{B_Q}{\rm 10^5~GeV} \right)^{3/2}
  \left( \frac{v}{\rm 10^{10}~GeV} \right)^{1/4}
  \left( \frac{\phi_0}{M_*} \right)^{2},
 \label{Omega_adec}
 \end{eqnarray}
 where we used $\xi\sin\theta_0\sim 1$.

Now, we can calculate the numerical value of the density parameter,
and see if it is cosmologically viable. For this purpose, we first
have to specify the dominant decay mode of the $R$-axion, since the
result depends on the decay width $\Gamma_a$. If the $R$-axion mass is
large enough (probably, for $m_a\gtrsim 1~{\rm GeV}$), the $R$-axion
decays into the gluon pair. In this case, we can use the calculation
based on the perturbative QCD, and the decay rate is given in the
formula in Eq.~(\ref{Gamma_a}) with $\alpha_{\rm SM} =\alpha_{\rm s}$.
However, if the $R$-axion mass becomes as small as (or smaller than)
$\sim 1~{\rm GeV}$, Eq.~(\ref{Gamma_a}) may not be reliable, since in
this case, decay rate into multi-meson final states has to be
calculated.  However, if the $R$-axion mass is light enough, decay
modes into multi-meson final states are kinematically forbidden.
Since the $R$-axion is a CP-odd particle, its mass has to be larger
than at least $3m_\pi$ for the decay into final states without
electromagnetic particles. Therefore, if $m_a<3m_\pi$,
$a\rightarrow\gamma\gamma$ is expected to be the dominant decay mode.
In this case, we can use the Eq.~(\ref{Gamma_a}) again with
$\alpha_{\rm SM} =\alpha_{\rm em}$. In the case $3m_\pi\leq
m_a\lesssim 1~{\rm GeV}$, estimation of the decay rate is quite
difficult, and we will not discuss this case further in this paper.
 
\begin{figure}
 \centerline{\epsfxsize=0.5\textwidth\epsfbox{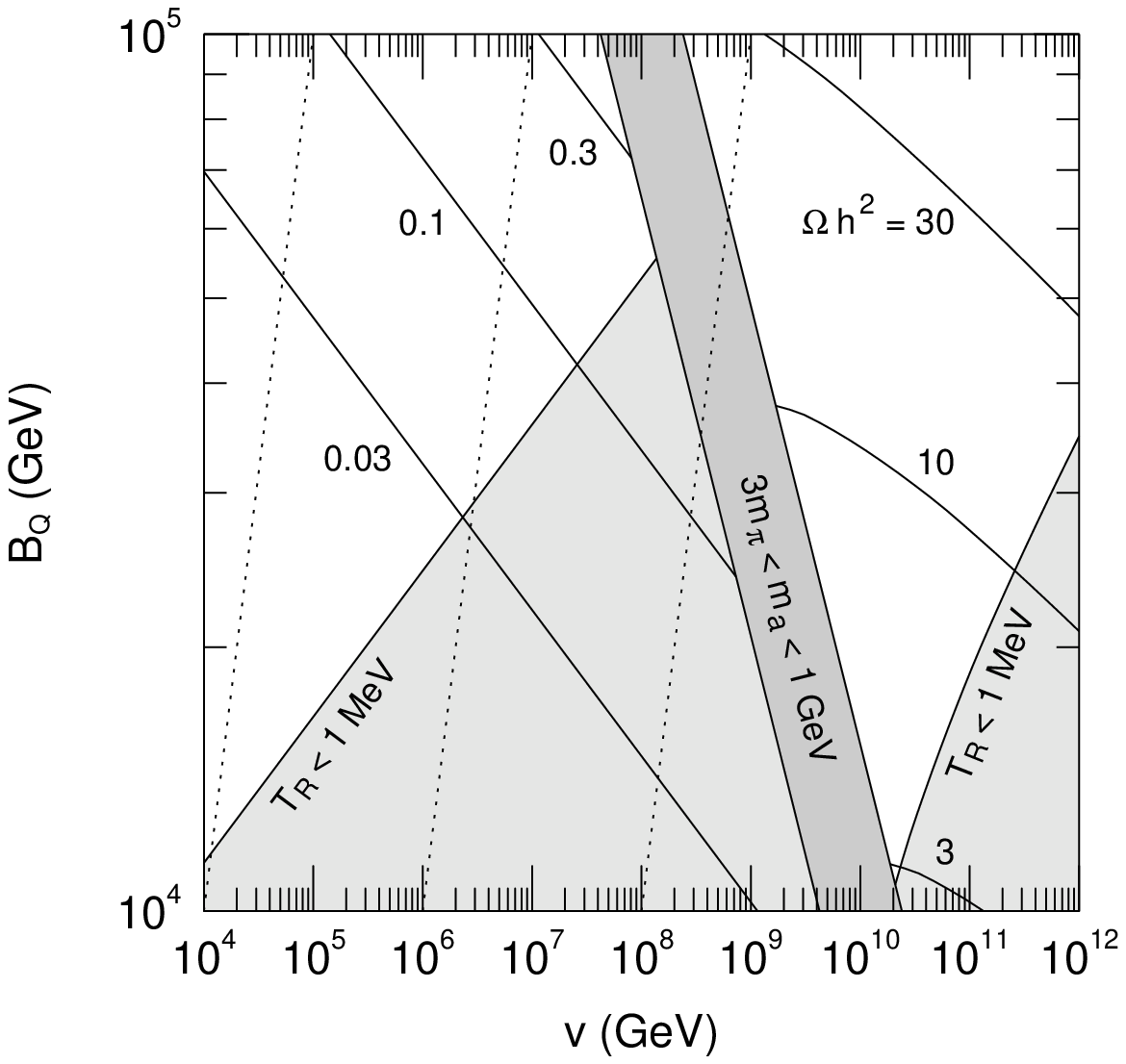}}
 \caption{Constant $\Omega_\phi h^2$ contours on the $v$ vs.~$B_Q$
plane with $\phi_0 =M_*$, $N_5=2$ and $k_0=1$. Region with $T_{\rm
R}<{\rm 1~MeV}$ is lightly shaded, and darkly shaded region
corresponds to $3m_\pi\leq m_a\leq {\rm 1~GeV}$.  Contours of constant
$v/F_X^{1/2}$ are also shown in dotted lines (1, 10, and 100, from
left to right).}
 \label{fig:Omg_k1}
 \vskip 1cm
 \centerline{\epsfxsize=0.5\textwidth\epsfbox{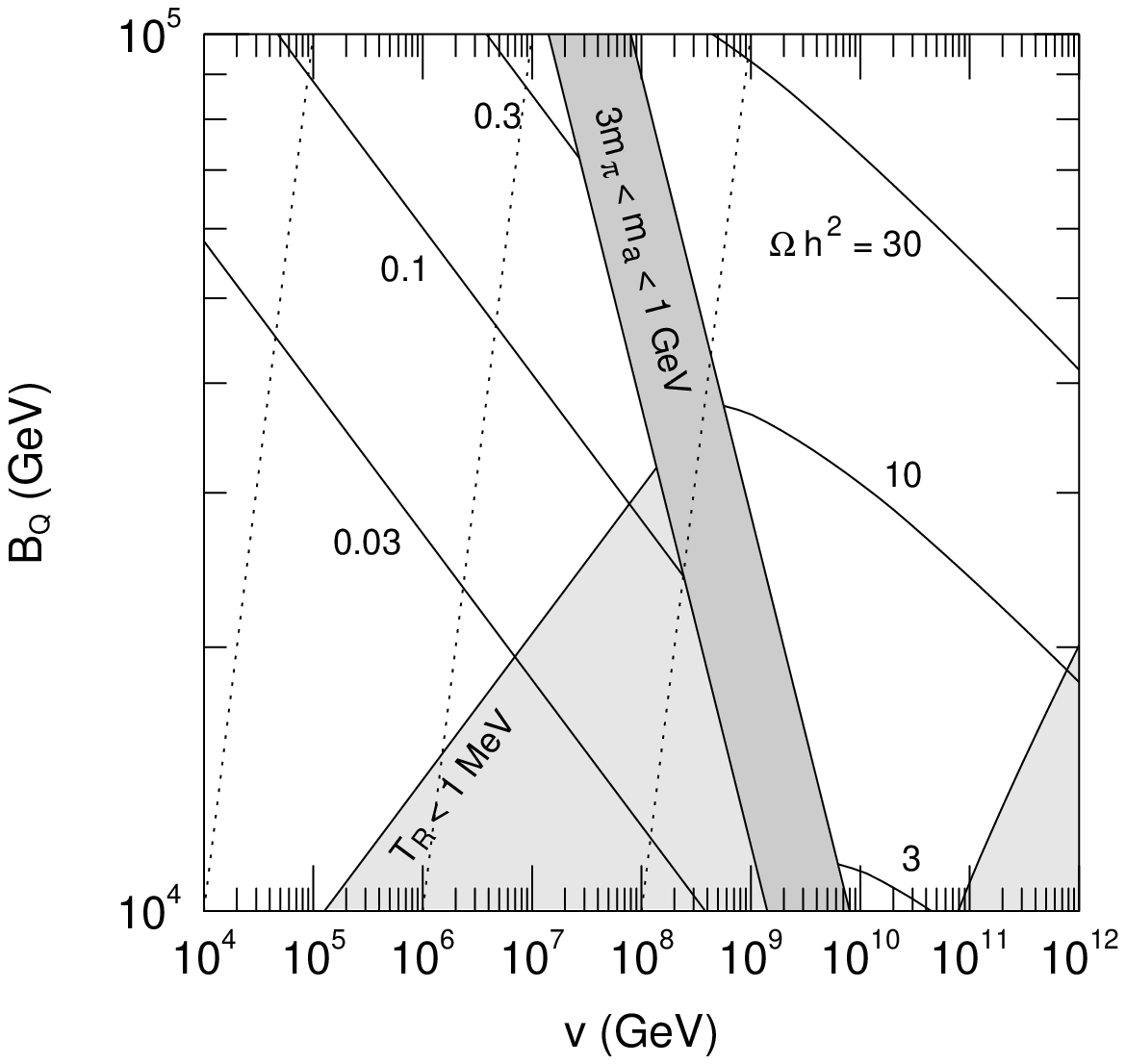}}
 \caption{Same as Fig.~\protect\ref{fig:Omg_k1}, except for $k_0=3$.}
 \label{fig:Omg_k3}
 \end{figure}

In Figs.~\ref{fig:Omg_k1} and \ref{fig:Omg_k3}, we plotted
$\Omega_\phi h^2$ on the $v$ vs.~$B_Q$ plane with $\phi_0=M_*$ and
$N_5=2$. In the calculation, we used Eq.~(\ref{rho2s_a}) with
$\xi\sin\theta_0=1$, and assumed that $a\rightarrow\gamma\gamma$ and
$a\rightarrow gg$ are the dominant decay modes of the $R$-axion for
$m_a\leq 3m_\pi$ and $m_a\geq {\rm 1~GeV}$, respectively. In these
figures, we shaded the region where the reheating temperature becomes
lower than 1~MeV. We also show the contours of the constant
$v/F_X^{1/2}$, which has to be larger than $\sim 1-10$.

Let us first discuss the case of $m_a\leq 3m_\pi$, where the $R$-axion
decays into the photon pair. This is the case for the smaller value of
$v$ (i.e., $v\lesssim 10^8-10^9{\rm ~GeV}$).  If the $R$-axion decays
into the photon pair, the reheating temperature becomes relatively
low.  However, if $B_Q$ is large enough, there is still a parameter
region where the reheating temperature is high enough ($T_{\rm
R}\gtrsim 1{\rm ~MeV}$). At the same time, $\Omega_\phi$ can be of
$O(0.1)$ or smaller, and hence the moduli field can be diluted enough
by the decay of the $R$-axion.

One problem of this case may be the non-perturbative effect discussed
in Section~\ref{sec:model}. In the case of $v\lesssim {\rm 10^9~GeV}$,
non-perturbative effect becomes sizable in the K\"ahler potential, and
hence the potential of $X$ cannot be well understood.  Therefore, we
need to make a dynamical assumption so that the SUSY breaking vacuum
exists. However, since the non-perturbative effect respects the
$R$-symmetry, important properties of the $R$-axion are unchanged.
Therefore, once we assume the existence of the SUSY breaking minimum,
our arguments are unchanged. One may also worry about the stability of
the vacuum if there is a SUSY preserving true minimum at the origin.
However, $T_{\rm R}\gtrsim {\rm 1~MeV}$ can be realized even with
$v/F_X^{1/2}\gtrsim 10$. Furthermore, this problem can be evaded in
models without SUSY preserving minimum.

For a larger value of $v$ (i.e., $v\gtrsim 10^9-10^{10}{\rm ~GeV}$),
the $R$-axion becomes heavier than $\sim {\rm 1~GeV}$, and decays into
the gluon pair. As we can see in Figs.~\ref{fig:Omg_k1} and
\ref{fig:Omg_k3}, if the $R$-axion decays into the gluons, the density
parameter of the moduli field becomes relatively larger. This is
because the decay rate is more enhanced, resulting in higher reheating
temperature. Even in this case, however, $\Omega_\phi\sim 10$ is
possible for $\phi_0\sim M_*$. Reduction of $\Omega_\phi$ by a factor
of about 10 may not be a serious problem.  For example, suppression of
the initial amplitude of the moduli field by a factor of 3 or so is
enough for $\Omega_\phi\leq 1$, since $\Omega_\phi$ is proportional to
$\phi_0^2$. We can naturally imagine an accidental cancellation of
this level. Notice that the non-perturbative effect on the K\"ahler
potential is not important in (most of) this case.

\subsection{Effect of the Real Part}

Next, we discuss the effect of the decay of the real part $\sigma$.
For this purpose, it is convenient to use the relation $m_{\rm
eff}\varphi^2\propto R^{-3}$ (with $\varphi =X$ and $\phi$).
Comparing this quantity at $|X|\sim X_0$ and $|X|\sim v$, we obtain
 \begin{eqnarray}
  \frac{m_{3/2}\phi^2}{m_\sigma\sigma^2} \sim 
  \left(\frac{\phi_0}{X_0}\right)^2.
 \label{m*phi^2}
 \end{eqnarray}
 The important point is that the effective mass for the moduli field
is always constant of $O(m_{3/2})$, while that for $X$ varies from
$\sim m_{3/2}$ to $m_\sigma$ which is of the order of the SSM scale.

Once $X$ gets trapped in the SUSY breaking minimum, $\sigma$ obeys the
parabolic potential, and hence Eq.~(\ref{m*phi^2}) leads
 \begin{eqnarray}
  \frac{\rho_\phi}{\rho_\sigma} \sim
  \frac{m_{3/2}}{m_\sigma} \left(\frac{\phi_0}{X_0}\right)^2,
 \label{ratiorho_s}
 \end{eqnarray}
 where $\rho_\sigma$ is the energy density of $\sigma$.

When the expansion rate becomes comparable to $\Gamma_\sigma$,
$\sigma$ decays. As discussed in Section~\ref{sec:model}, $\sigma$
dominantly decays into the $R$-axion pair. The number density of
$\sigma$, $n_\sigma$, is related to the energy density as
$\rho_\sigma=m_\sigma n_\sigma$. Therefore, by using
Eq.~(\ref{ratiorho_s}) with $|X_0|\sim M_*$, we obtain
 \begin{eqnarray}
  \frac{\rho_\phi}{n_a^{\rm dec}} \sim
  m_{3/2} \left(\frac{\phi_0}{M_*}\right)^2,
 \label{rhophi/nadec}
 \end{eqnarray}
 where $n_a^{\rm dec}$ is the number density of the $R$-axion produced
by the decay of $\sigma$. Notice that this ratio is constant even with
entropy production.

As the Universe expands, the emitted $R$-axions are red-shifted and
eventually become non-relativistic. Then, when they decay, they also
contribute to the entropy production to dilute the moduli field.
Comparing Eq.~(\ref{rhophi/nadec}) with Eq.~(\ref{phophi/nr}), we see
that the number of the $R$-axion produced by the decay of $\sigma$ is
comparable to that of the coherent mode. Therefore, the dilution by
this incoherent $R$-axion is of the same order of that by the coherent
mode, and the results given in the previous subsection are almost
unchanged.

Finally, we discuss the effect of the decay mode into the gravitino
pair. As discussed in Section~\ref{sec:model}, $\sigma$ and $a$ may
decay into the gravitino. Since the gravitino is stable in GMSB, the
mass density of the gravitino, $\rho_{3/2}$, should not exceed the
closure limit.

Let us consider the gravitino production due to the decay of $\sigma$
as an example. In order to discuss the mass density of the gravitino
by the decay of $\sigma$ (and of the $R$-axion), it is convenient to
consider the ratio $m_{3/2}n_{3/2}/\rho_\phi\sim\rho_{3/2}/\rho_\phi$,
where $n_{3/2}$ is the number density of the gravitino.  This ratio
can be easily calculated from Eq.~(\ref{ratiorho_s}). When $\sigma$
decays, the number density of the gravitino is given by
 \begin{eqnarray}
  n_{3/2} 
  \sim 
  {\rm Br} (\sigma\rightarrow\psi_\mu\psi_\mu) \times n_{\sigma}
  \sim 
  {\rm Br} (\sigma\rightarrow\psi_\mu\psi_\mu)
  \times \frac{\rho_\sigma}{m_\sigma}.
 \end{eqnarray}
 Combining this equation with Eq.~(\ref{ratiorho_s}) and $|X_0|\sim
M_*$, we obtain
 \begin{eqnarray}
  \frac{m_{3/2}n_{3/2}}{\rho_\phi}
  \sim
  {\rm Br} (\sigma\rightarrow\psi_\mu\psi_\mu)
  \times \left(\frac{\phi_0}{M_*}\right)^{-2}.
 \end{eqnarray}
 Importantly, this ratio remains constant even after a large entropy
production. Once the gravitinos are red-shifted to be
non-relativistic, the above ratio becomes $\rho_{3/2}/\rho_\phi$.  As
a result, the density parameter of the gravitino, $\Omega_{3/2}$, is
related to $\Omega_\phi$ as
 \begin{eqnarray}
  \Omega_{3/2}
  \sim 
  {\rm Br} (\sigma\rightarrow\psi_\mu\psi_\mu)\times
  \Omega_\phi
  \left(\frac{\phi_0}{M_*}\right)^{-2}.
 \end{eqnarray}
 It is also straightforward to check that the density parameter of the
gravitino from the decay of the $R$-axion is given by a similar
formula (with extra coefficient of $O(1)$) with the relevant branching
ratio for the $R$-axion.

With the branching ratio given in Eqs.~(\ref{Br_a}) and (\ref{Br_s}),
we can see that the mass density of the gravitino is small enough. By
using the formula for $m_\sigma$, ${\rm Br}
(\sigma\rightarrow\psi_\mu\psi_\mu)$ is estimated to be of
$O(10^{-4})$.  Furthermore, in most of the parameter region we are
interested in, the $R$-axion mass is smaller than about 10~GeV, and
hence ${\rm Br} (a\rightarrow\psi_\mu\psi_\mu)$ is at most of
$O(10^{-2})$. Therefore, even with $\Omega_\phi\times
(\phi_0/M_*)^{-2}\sim 10$ for $v\gtrsim 10^9-10^{10}{\rm ~GeV}$ (i.e.,
for $m_a\gtrsim {\rm 1~GeV}$), the energy density of the gravitino is
small enough.

\subsection{Baryogenesis and the Flat Direction in SSM}

So far, we have discussed the dilution of the moduli field by the
decay of the $R$-axion. In this scenario, one may worry about the
baryogenesis, since the reheating temperature seems to be too low to
generate the baryon asymmetry. In this subsection, we briefly comment
that enough baryon asymmetry can be generated by the Affleck-Dine
mechanism~\cite{NPB249-361}. (For a detailed discussion, see
Ref.~\cite{PRD56-1281}.)

Important assumptions for the Affleck-Dine baryogenesis are that SSM
flat direction $\varphi_{\rm SSM}$ (for Affleck-Dine baryogenesis, we
call it Affleck-Dine field) has a large amplitude in the early
Universe, and that there is baryon-number breaking term in the
potential of $\varphi_{\rm SSM}$. With these assumptions,
non-vanishing baryon-number can be generated as the Affleck-Dine field
evolves. Physics in this mechanism is basically the same as that in
the case of the $R$-number generation.\footnote
 {The structures of the baryon-number breaking terms may be different.
However, it does not affect the following discussions, as far as the
initial amplitude of $\varphi_{\rm SSM}$ is of $O(M_*)$.}
 Since the potential for the Affleck-Dine field is dominated by the
supergravity contribution for $\varphi_{\rm SSM}\sim M_*$, the ratio
of the energy density of the moduli field to the baryon number density
$n_{\rm B}$ is estimated as~\cite{PRD56-1281}
 \begin{eqnarray}
  \frac{\rho_\phi}{n_{\rm B}} \sim m_{3/2}
  \left(\frac{\phi_0}{M_*}\right)^2,
 \label{rhophi/nB}
 \end{eqnarray}
 where we applied a similar argument as in the $R$-axion case. Here,
we assumed that the initial amplitude of $\varphi_{\rm SSM}$ is of
$O(M_*)$ and that source of CP violation is of $O(1)$, which are
corresponding to $\xi\sim 1$ and $\sin\theta_0\sim 1$ in the $R$-axion
case, respectively. Notice that this ratio is constant after the
moduli and Affleck-Dine fields start to move.

With Eq.~(\ref{rhophi/nB}), we can estimate the baryon number density
of the present Universe, once we fix the current mass density of the
moduli field. As we discussed in the previous subsections, present
mass density of the moduli field depends on the magnitude of the
entropy production. In this subsection, we just assume the moduli
field is diluted enough by the decay of the $R$-axion. Then, adopting
$\Omega_\phi\lesssim 1$, baryon-to-entropy ratio is estimated
as~\cite{PRD56-1281}
 \begin{eqnarray}
  \frac{n_{\rm B}}{s} \lesssim 4\times 10^{-5}
  \times h^{-2}
  \left( \frac{m_{3/2}}{\rm 100~keV} \right)^{-1}
  \left(\frac{\phi_0}{M_*}\right)^{-2}.
 \label{nB/s}
 \end{eqnarray}
 Comparing the above relation with the baryon number density required
from the big-bang nucleosynthesis ($n_{\rm B}/s\sim
O(10^{-11})$~\cite{hph9805405}), we can see that enough baryon
asymmetry can remain even if there is a large entropy production to
dilute the moduli field.  In fact, as can be seen in Eq.~(\ref{nB/s}),
baryon-to-entropy ratio may be too large for some value of the
gravitino mass, if we adopt a naive initial condition for the
Affleck-Dine field. However, this problem may be solved by adopting a
smaller value of the source of CP violation, or a smaller value of the
initial amplitude of the Affleck-Dine field.  Therefore, we do not
worry about this issue.

Even apart from the baryogenesis, effect of the SSM flat direction may
be interesting, since it may produce a large entropy. The potential
for the flat direction has a similarity to that of the $\sigma$ field;
it is parabolic around the minimum (i.e., around the origin), then the
potential is lifted only logarithmically once the amplitude becomes
larger than the messenger scale, and finally, supergravity effect
dominates when the amplitude is close $M_*$.  Therefore, its effective
mass varies from $\sim m_{3/2}$ to the SSM scale, if $\varphi_{\rm
SSM}$ changes its amplitude from $\sim M_*$ to $\sim 0$. By applying
the similar argument as in the $\sigma$ case, we can estimate the
density parameter of the moduli field:
 \begin{eqnarray}
  \Omega_\phi h^2 \sim 2\times 10^3 \times
  \left(\frac{m_{\rm SSM}}{\rm 1~TeV}\right)^{-1}
  \left(\frac{m_{3/2}}{\rm 1~keV}\right)
  \left(\frac{T_{\rm SSM}}{\rm 10~TeV}\right)
  \left(\frac{\phi_0}{M_*}\right)^2,
 \end{eqnarray}
 where $T_{\rm SSM}$ is the reheating temperature due to the decay of
the flat direction. Comparing this result with Eq.~(\ref{Omg_noD}), we
can see that a large entropy production is possible even from the flat
direction in SSM.

However, usually, the reheating temperature is of $O(10~{\rm TeV})$,
and hence the density parameter becomes typically of $O(10^3)$ even
for $m_{3/2}\sim {\rm 1~keV}$. Therefore, it seems difficult to dilute
the moduli density enough in a simple scenario, unless we come up with
a model with very light gravitino. However, if the reheating
temperature can be somehow lowered down to $\sim 10~{\rm GeV}$,
dilution due to the decay of the SSM flat direction may be enough to
solve the cosmological moduli problem. For example, Pauli blocking may
delay the decay of the flat direction, as pointed out in
Ref.~\cite{PLB328-354}, although the baryon asymmetry may be generated
too much if we naively apply the argument in Ref.~\cite{PLB328-354}.

\section{Discussion} 
 \label{sec:discussion} 
 \setcounter{equation}{0}

In this paper, we have discussed the cosmology based on the direct
gauge mediation model with the inverted hierarchy mechanism. In
particular, we have studied the implication of the SUSY breaking field
on the cosmology of GMSB.

If the SUSY breaking flat direction $X$ initially has a very large
amplitude of $O(M_*)$, it can be a source of the large entropy
production. In particular, once the amplitude of $X$ becomes a few
orders of magnitude smaller than $M_*$, potential for $X$ is dominated
by the logarithmic piece. Then, the energy density of $X$ decreases
more slowly than those of the scalars with quadratic potential, and
the SUSY breaking field may play a very important role in cosmology.

In particular, the entropy production by the decay of the $R$-axion
may be so large that the energy density of the moduli field can be
diluted enough.  Therefore, the direct gauge mediation models with the
inverted hierarchy mechanism contain a natural candidate of the large
entropy production to solve the serious cosmological moduli
problem. We have also seen that enough baryon asymmetry can be
generated by the Affleck-Dine mechanism even with this large entropy
production.

In our discussion, we mainly focused on the case of the direct gauge
mediation model with the inverted hierarchy mechanism. However, the
scenario discussed in this paper can be applied to a larger class of
models, since the most important building block is just the
logarithmically lifted potential of the SUSY breaking field for
$|X|\gtrsim v$.  Consequently, if a direct gauge mediation model uses
the mechanism of SUSY breaking proposed by Izawa and Yanagida, and by
Intriligator and Thomas~\cite{IzaYanInt}, it automatically contains a
reasonable source of large entropy production.  Of course, it is
non-trivial for such a model to stabilize the potential of the SUSY
breaking field.  The inverted hierarchy mechanism provides one
attractive mechanism for the
stabilization~\cite{PRL79-18,NPB510-12,PLB413-336,hph9804450}.  Other
approach may be to use a non-perturbative effect on the K\"ahler
potential~\cite{nonpert}.

One may worry about the astrophysical constraints on the $R$-axion, as
in the case of the QCD axion.  However, since the $R$-axion has a
larger mass than the QCD axion, constraints on the $R$-axion are much
weaker.  Constraints from the cooling of the horizontal-branch (HB)
stars~\cite{HBstar} are evaded, since we consider $R$-axion heavier
than 100~keV (see Eq.~(\ref{ma})) while the core temperature the HB
stars are typically of $O({\rm 10~keV})$. Furthermore, the $R$-axion
does not affect the background UV light~\cite{UVlight}, since it has
already decayed away.  Constraints from SN1987A are more non-trivial.
In order not to affect the cooling of SN1987A, QCD axion with the
decay constant from $\sim{\rm 10^6~GeV}$ to $\sim{\rm 10^9~GeV}$ is
forbidden~\cite{SN1987A}. However, for this range of $v$, the
$R$-axion mass is (almost) always larger than the core temperature of
SN1987A (i.e., $O({\rm 10~MeV})$), and hence the emission of the
$R$-axion is suppressed.  (However, small parameter space around
$v\sim{\rm 10^6~GeV}$ and $B_Q\sim {\rm 10^4~GeV}$ may be excluded,
though the reheating temperature is too low in this region.)  When
$v\lesssim{\rm 10^6~GeV}$, the $R$-axion is thermalized enough in
SN1987A, and it does not affect the cooling process. The QCD axion
with small decay constant may be detected in water \v{C}erenkov
detectors~\cite{PRL65-960}. However, the $R$-axion cannot be
constrained with this method, since $R$-axion decays before reaching
the earth.  This fact suggests another constraint on the $R$-axion; if
the emitted $R$-axions decay into the photons on the way to the earth,
apparent luminosity of SN1987A may be increased. Therefore, for our
scenario, light $R$-axion is potentially dangerous. However, the
estimation of the $R$-axion flux is very complicated, in particular
since the $R$-axion may have a mass comparable to the core temperature
of SN1987A. Therefore, it is an open question which parameter region
is excluded from this argument. Notice that, if the $R$-axion mass is
heavier than of $O({\rm 10~MeV})$, this problem can be evaded thanks
to the Boltzmann suppression.

In this paper, we have not paid attention to the primordial inflation,
since it is beyond the scope of this paper. Of course, it is important
to find a viable candidate of the inflaton for the primordial
inflation, and some efforts are made in this issue~\cite{PRL79-2632}.
Here, we just mention that, in our scenario, inflaton for the
primordial inflation does not have to decay into the particles in the
SSM sector. Since the background radiation and baryons in the present
Universe originate to the decay of the SUSY breaking field $X$ (and
probably, to the decay of the Affleck-Dine field), primordial
inflation is not required to reheat the SSM sector.  In an extreme
case, inflaton may decay only into particles in the hidden sector.
Even if the energy density of the Universe is once dominated by that
of the hidden sector particles, it is eventually diluted by the
entropy production by the decay of the SUSY breaking field. This fact
relaxes the conventional requirements on the inflaton which is usually
required to decay into the SSM particles.

It is interesting to consider candidates of the CDM in this scenario.
Because of the low reheating temperature after the decay of the
$R$-axion, thermal productions of any known candidates are
inefficient. In this case, a possible candidate is the coherent
oscillation of the moduli field. Indeed, if the energy density of the
moduli field can be diluted to be $\Omega_\phi\sim 1$ by the decay of
the $R$-axion, they can be a viable candidate of the CDM.  This
scenario is constrained by the line spectrum of the background cosmic
$X$-ray emitted from the decay of the moduli field~\cite{Xray}.  Even
if the lifetime is much longer than the age of the Universe, some
fraction of the moduli field has already decayed into the photons, and
it contributes to the background $X$-ray spectrum.  Due to the
negative observation of such a line spectrum, moduli field heavier
than about 200~keV is forbidden, if $\Omega_\phi
h^2=1$~\cite{Xray}.\footnote
 {However, if the interaction of the moduli field with the photon has
extra suppression, lifetime of the moduli field gets longer and this
constraint becomes less stringent. In this case, heavier moduli field
is still allowed. For example, this may happen if the
moduli-photon-photon vertex is induced by loop effects.}
 In the direct gauge mediation model, the gravitino mass (i.e., the
moduli mass) can be lighter than about 100~keV, and hence the moduli
can be a good candidate of the CDM.

Cosmological implication of the $R$-axion in other classes of GMSB is
another interesting issue. In general, dynamical SUSY breaking
requires spontaneously broken $R$-symmetry~\cite{NPB416-46}.
Therefore, all the dynamical SUSY breaking models contain $R$-axion in
the low energy spectrum. Then, there is a possibility of large
$R$-number generation by the mechanism we discussed, if the SUSY
breaking field has a large initial amplitude. However, the interaction
of the $R$-axion is model-dependent, and in some case, its decay rate
may be much more suppressed. In this case, it causes a cosmological
difficulty, since the reheating temperature after the decay of the
$R$-axion becomes too low for the big-bang nucleosynthesis. Of course,
this problem itself can be always evaded by assuming a small initial
amplitude of the SUSY breaking field.

Since different models introduce different sets of new particles which
have various properties, detailed cosmological scenario is
model-dependent. Therefore, one should always keep in mind that the
SUSY breaking field (and all the new degrees of freedom) may change
the conventional arguments on the cosmology based on supersymmetric
models. In some case, it may cause a serious cosmological disaster,
but in other case, as we have seen, it may provide a natural and
well-motivated solution to several serious cosmological difficulties.

\section*{Acknowledgement}

The author would like to thank G.~Giudice, M.~Kawasaki, H.~Murayama,
J.~Terning, and especially K.~Agashe for useful discussions.

This work was supported by the Director, Office of Energy Research,
Office of High Energy and Nuclear Physics, Division of High Energy
Physics of the U.S. Department of Energy under Contract
DE-AC03-76SF00098.

\appendix

\section{Example of the Model}
 \label{app:example}
 \setcounter{equation}{0}

In this Appendix, we show an example of the direct gauge mediation
model in which vacuum is stabilized by the inverted hierarchy
mechanism. The model, which is originally proposed in
Ref.~\cite{NPB510-12}, is based on the symmetry ${\rm SU(2)_{B1}\times
SU(2)_{B2}\times SU(2)_S\times G_{SM}}$. In this model, SU(2)$_{\rm
B1}$ is a gauge group which stabilize the minimum of the potential,
while SU(2)$_{\rm S}$ provides a strong gauge interaction which
induces gaugino condensation to break SUSY.  Furthermore, G$_{\rm SM}$
is the standard model gauge group. The particle content is shown in
the Table~\ref{table:example}.

Superpotential in this model is given by
 \begin{eqnarray}
  W = y_Q\Sigma\bar{Q}Q 
  + y_3\Sigma\bar{q}_3q_3 + y_2\Sigma\bar{q}_2q_2 
  + y_1\Sigma\bar{q}_1q_1.
 \end{eqnarray}
 With this superpotential, we concentrate on the flat direction of
$X\sim ({\rm det}\Sigma)^{1/2}$. Along $X$, we parametrize
 \begin{eqnarray}
  \Sigma = \frac{1}{\sqrt{2}} {\rm diag} (X,X).
 \label{Sigma_flat}
 \end{eqnarray}
 Once $\Sigma$ gets this VEV, $Q$ and $\bar{Q}$ acquire a mass of
$m_Q\simeq \frac{1}{\sqrt{2}}y_QX$. For $\mu\lesssim m_Q$, SU(2)$_{\rm
S}$ is a pure SUSY Yang-Milles theory, and gaugino condensation
induces the superpotential of the form $W_{\rm eff}=2\Lambda_{\rm
eff}^3$, where $\Lambda_{\rm eff}$ is the strong scale of SU(2)$_{\rm
S}$ below the mass scale of $Q$ and $\bar{Q}$. By matching the strong
scales for the theory below and above $m_Q$, we obtain
 \begin{eqnarray}
  W_{\rm eff} = \sqrt{2} y_Q X \Lambda^2,
 \end{eqnarray}
 where $\Lambda$ is the strong scale for the theory above $m_Q$. Since
the superpotential is linear in $X$, $F$-component of $X$ has a VEV of
$F_X=\sqrt{2}y_Q\Lambda^2$, and SUSY is broken.

The minimum of the potential is determined by the inverted hierarchy
mechanism~\cite{PLB105-267}. At the tree level, potential for $X$ is
completely flat, and hence the scale dependence of the wave function
normalization of $\Sigma$ determines the position of the minimum. In
this case, the potential for $X$ is given by
 \begin{eqnarray}
  V = \frac{F_X^2}{Z_\Sigma (X^*,X)},
 \end{eqnarray}
 where $Z_\Sigma (X^*,X)$ is the wave function normalization of
$\Sigma$ which is evaluated at $\mu =|X|$. At the 1-loop level, RGE
for $Z_\Sigma$ is given by
 \begin{eqnarray}
  \frac{d\ln Z_\Sigma}{dt} = \frac{1}{16\pi^2}
  \left( \frac{3}{2}g_{\rm B1}^2 +\frac{3}{2}g_{\rm B2}^2 
  - 2 y_Q^2 - 3 y_3^2 - 2 y_2^2 - y_1^2 \right),
 \label{2225_rge}
 \end{eqnarray}
 where $g_{\rm B1}$ and $g_{\rm B2}$ are the gauge coupling constants
for SU(2)$_{\rm B1}$ and SU(2)$_{\rm B2}$, respectively.  Thus, $X$
has an extremum at $X=v$, where
 \begin{eqnarray}
  \frac{3}{2}(g_{\rm B1}^2+g_{\rm B2}^2) = 
  2y_Q^2 + 3y_3^2 + 2y_2^2 + y_1^2.
 \label{2225_vac}
 \end{eqnarray}

 \begin{table}[t]
 \begin{center}
 \begin{tabular}{ccccccc}
 \hline\hline
  {} & {SU(2)$_{\rm B1}$} & {SU(2)$_{\rm B2}$} & {SU(2)$_{\rm S}$} &
  {SU(3)$_{\rm C}$} & {SU(2)$_{\rm L}$} & {U(1)$_{\rm Y}$}
 \\ \hline
  {$\Sigma$} & {${\bf 2}$} & {${\bf 2}$} & {${\bf 1}$} & 
  {${\bf 1}$} & {${\bf 1}$} & {$0$}
 \\
  {$Q$} & {${\bf 2}$} & {${\bf 1}$} & {${\bf 2}$} & 
  {${\bf 1}$} & {${\bf 1}$} & {$0$}
 \\
  {$\bar{Q}$} & {${\bf 1}$} & {${\bf 2}$} & {${\bf 2}$} & 
  {${\bf 1}$} & {${\bf 1}$} & {$0$}
 \\
  {$q_3$} & {${\bf 2}$} & {${\bf 1}$} & {${\bf 1}$} & 
  {${\bf 3}$} & {${\bf 1}$} & {$-1/3$}
 \\
  {$\bar{q}_3$} & {${\bf 1}$} & {${\bf 2}$} & {${\bf 1}$} & 
  {${\bf 3^*}$} & {${\bf 1}$} & {$1/3$}
 \\
  {$q_2$} & {${\bf 2}$} & {${\bf 1}$} & {${\bf 1}$} & 
  {${\bf 1}$} & {${\bf 2}$} & {$1/2$}
 \\
  {$\bar{q}_2$} & {${\bf 1}$} & {${\bf 2}$} & {${\bf 1}$} & 
  {${\bf 1}$} & {${\bf 2}$} & {$-1/2$}
 \\
  {$q_1$} & {${\bf 2}$} & {${\bf 1}$} & {${\bf 1}$} & 
  {${\bf 1}$} & {${\bf 1}$} & {$0$}
 \\
  {$\bar{q}_1$} & {${\bf 1}$} & {${\bf 2}$} & {${\bf 1}$} & 
  {${\bf 1}$} & {${\bf 1}$} & {$0$}
 \\ \hline\hline
 \end{tabular}
 \caption{Particle content of a direct gauge mediation model given in
Ref.~\protect\cite{NPB510-12}.}
 \label{table:example}
 \end{center} 
 \end{table}

In order to see whether this is a minimum or a maximum, it is
convenient to estimate the mass of the real part of $X$, which we call
$\sigma$.\footnote
 {Imaginary part is pseudo-Nambu-Goldstone boson, and its mass is from
the supergravity effect, as discussed in Section~\ref{sec:model}.}
 For simplicity, we consider the case where $g_{\rm B2}$, $y_3$, and
$y_2$ are small. (Even in the general case, the following discussion
is qualitatively correct.)  Then, mass of $\sigma$ is given by
 \begin{eqnarray}
  m_\sigma^2
  \simeq \frac{1}{(16\pi^2)^2}
  \left(
  \frac{33}{4} g_{\rm B1}^4 + 24 y_Q^4
  - 24 g_{\rm B1}^2 y_Q^2 - 6 g_{\rm S}^2 y_Q^2
  \right) \frac{F_X^2}{v^2},
 \label{2225_msigma}
 \end{eqnarray}
 where $g_{\rm S}$ is the gauge coupling constant of SU(2)$_{\rm S}$.
Importantly, $g_{\rm S}$ is usually large in order to induces the
strong dynamics to break SUSY. Then, for large $y_Q$, $g_{\rm
S}^2y_Q^2$ term becomes so large that $m_\sigma^2$ becomes negative.
(For large enough $y_Q$, $m_\sigma^2$ may become positive, but $y_Q$
blows up below the Planck scale.) Numerically, $y_Q$ cannot be larger
than 0.2 -- 0.3 for the positivity of $m_\sigma^2$. For $v\gtrsim
10^9{\rm ~GeV}$, solution to Eq.~(\ref{2225_vac}) with positive
$m_\sigma^2$ can be found with reasonable values of the coupling
constants.

For $v\lesssim 10^9{\rm ~GeV}$, we cannot neglect the non-perturbative
effects, as we discussed in Section~\ref{sec:model}. In this case,
K\"ahler potential is dominated by the non-perturbative piece, and it
is unclear whether there can be a minimum. However, since the
non-perturbative effects are not well understood, there is a
possibility to have a stable minimum even with the non-perturbative
effect. Therefore, for $v\lesssim 10^9{\rm ~GeV}$, we make a
dynamical assumption so that the stable SUSY breaking vacuum exists.
Notice that, in this case, upper bound on $y_Q$ is irrelevant.

Once the SUSY is broken and the VEV of $X$ is fixed, SUSY breaking is
mediated down to the SSM sector by integrating out messengers, $q_i$
and $\bar{q}_i$ ($i=3,2$). Since these are the only superfields with
standard model quantum numbers which couple to the SUSY breaking
field, SUSY breaking masses obey the well-known mass
formula~\cite{NPB207-96}.  Notice that, in this model, $N_5=2$ (see
Eq.~(\ref{Gamma_a})).
 
Several remarks are in order. First, for $\mu\lesssim v$, diagonal
SU(2) remains if SU(2)$_{\rm B2}$ is gauged. For our cosmological
scenario, this may be dangerous, since the $R$-axion also couples to
this gauge field. If $R$-axion dominantly decays into this gauge
field, energy density of the standard model particles cannot be
generated enough since the gauge field for this diagonal SU(2) does
not couple to the SSM sector. This problem can be evaded if
SU(2)$_{\rm B2}$ interaction is weak enough (i.e., much weaker than
the electromagnetic interaction, or maybe not gauged). Another comment
is on the true vacuum of the potential. In fact, this model has a SUSY
preserving true vacuum at the origin ($|X|=0$).  However, tunneling
rate from the SUSY breaking minimum to the true minimum is small
enough, if the SUSY breaking vacuum is far enough away from the
confinement scale $\Lambda$; numerically, $v/\Lambda\gtrsim
10$~\cite{NPB510-12}.

\section{Shift of the Minimum of the Moduli Potential}
 \label{app:shift}
 \setcounter{equation}{0}

In this Appendix, we consider how the minimum of the moduli potential
shifts when the SUSY breaking field $X$ is oscillating.  When the
Universe is dominated by $X$, extra terms are induced in the moduli
potential by supergravity effects, and they may change the minimum of
the moduli potential. If this shift is too large, it may change our
argument, since the moduli field may oscillate around a shifted
minimum.  In this Appendix, we see that this effect is not significant
for our case, and that our naive calculations are relevant.

First, let us consider possible modifications of the moduli potential
in the presence of $X$. In supergravity, there can be two effects. One
is from the non-vanishing VEV of $X$; since $X$ and $\phi$ may have
Planck-suppressed interactions, potential may have terms which are
proportional to the powers of $(|X|/M_*)$. The other is from the
expansion rate $H$ induced by the condensation of $X$; since the
scalar potential contains a term of the form $\sim e^{K/M_*^2}V$,
non-vanishing potential energy induces terms proportional to
$H^2$~\cite{DinRanTho}.

With these effects, linear term is induced in the moduli potential,
which shifts the minimum of the potential:
 \begin{eqnarray}
  V(\phi)\sim m_{3/2}^2 \phi^* \phi
  - m_{3/2}^2 (\bar{\phi}^* \phi + {\rm h.c.}),
 \label{V(phi)_XH}
 \end{eqnarray}
 where we define the origin of the moduli field so that VEV of $\phi$
vanishes for the empty background. Here, the second term in
Eq.~(\ref{V(phi)_XH}) is the induced term, and $\bar{\phi}$ is given
by
 \begin{eqnarray}
  \bar{\phi} \sim {\rm max} \left[|X|, (H^2/m_{3/2}^2)M_* \right].
 \end{eqnarray}
 (In this section, we neglect $O(1)$ coefficients which do not change
our argument.) Notice that $\bar{\phi}$ is the shifted minimum of the
potential, and that it is time-dependent.\footnote
 {Other terms (higher order terms) are less significant for our
argument, and they do not change the following discussion. For
example, if the potential has a term of the form $H^2\phi^2$, solution
to the equation of motion contains a term of
$O(H^2/m_{3/2}^2)\phi_{\rm osc}$, with $\phi_{\rm osc}$ being the
solution to the equation of motion with $H=0$. However, this is much
smaller than the original amplitude $\phi_{\rm osc}$ since $H\ll
m_{3/2}$, and hence this effect is negligible.}
 With the above potential, equation of motion for $\phi$ is given by
 \begin{eqnarray}
  \ddot{\phi} + 3H\dot{\phi} 
  + m_{3/2}^2 (\phi-\bar{\phi})
  = 0.
 \label{eq4phi_XH}
 \end{eqnarray}
 Solution to this equation can be written as
 \begin{eqnarray}
  \phi = \phi_{\rm osc} + \delta \phi,
 \label{phiosc+delphi}
 \end{eqnarray}
 where $\phi_{\rm osc}$ is the oscillating solution with
$\bar{\phi}=0$, while $\delta\phi$ is a perturbation induced by the
new terms. Notice that $\phi_{\rm osc}$ obeys the original equation of
motion:
 \begin{eqnarray}
  \ddot{\phi}_{\rm osc} + 3H\dot{\phi}_{\rm osc}
  + m_{3/2}^2 \phi_{\rm osc}
  = 0,
 \end{eqnarray}
 and hence its averaged amplitude is proportional to $R^{-3/2}$. Thus,
$\phi_{\rm osc}$ obeys the behavior discussed in
Section~\ref{sec:evolution}.

If $\bar{\phi}\sim |X|$, the shift cannot be larger than the averaged
amplitude of $\phi_{\rm osc}$ (see Eq.~(\ref{rel_XvsPhi})). In this
case, the original amplitude is always larger than the shift of the
minimum, and hence the extra contribution is negligible. Thus, in the
following discussion, we concentrate on the case where $\bar{\phi}$ is
dominated by the Hubble-induced term.

In order to consider the Hubble-induced term, we approximate the
potential of $X$ as
 \begin{eqnarray}
  V\sim \frac{\zeta_2}{(16\pi^2)^2} m_{3/2}^2M_*^2
  \left( \ln\frac{X^*X}{v^2} \right)^2,
 \label{V(X)_approx}
 \end{eqnarray}
 with $\zeta_2$ being a constant. This potential has a minimum at
$|X|=v$, and increases logarithmically for large $|X|$. Therefore,
this potential reproduces the important feature of the potential of
$X$ (especially for $|X|\sim v$). When the Universe is dominated by
$X$, expansion rate is estimated as
 \begin{eqnarray}
  H \sim \frac{\sqrt{\zeta_2}}{16\pi^2} L m_{3/2},
 \label{H_log}
 \end{eqnarray}
 with 
 \begin{eqnarray}
  L \equiv \ln\frac{X^*X}{v^2}.
 \end{eqnarray}
 Importantly, this expansion rate is much smaller than the gravitino
mass. With the above expansion rate, we denote
 \begin{eqnarray}
  \bar{\phi} = \frac{k_HH^2}{m_{3/2}^2} M_*,
 \end{eqnarray}
 where $k_H$ is an unknown constant expected to be of $O(1)$. We
consider the case where the energy density of the Universe is
dominated by that of $X$, so $H$ decreases as the amplitude of $X$
approaches to $v$.

As shown in Eq.~(\ref{phiosc+delphi}), solution to
Eq.~(\ref{eq4phi_XH}) is given by the sum of the oscillating solution
$\phi_{\rm osc}$ which is from the non-perturbed equation of motion
and $\delta\phi$ induced by extra terms with non-vanishing $H$. We
have already understood the behavior of $\phi_{\rm osc}$, so now we
consider $\delta\phi$.

The important point in deriving $\delta\phi$ is that the expansion
rate $H$ is much smaller than the gravitino mass $m_{3/2}$. Because of
this, we can expand $\delta\phi$ by powers of $(H^2/m_{3/2}^2)$.
First, let us consider a simple case where $\dot{H}$ is proportional
to $H^2$:
 \begin{eqnarray}
  \dot{H} = - c_H H^2,
 \end{eqnarray}
 with $c_H$ being a constant of $O(1)$. For example, if $X$ has a
parabolic potential, $c_H=3/2$. In this case, solution to
Eq.~(\ref{eq4phi_XH}) is obtained as
 \begin{eqnarray}
  \phi = \phi_{\rm osc} + \left[
  \frac{k_HH^2}{m_{3/2}^2}
  - 6 c_H (c_H-1)
  \frac{k_HH^4}{m_{3/2}^4}
  + O \left(\frac{H^6}{m_{3/2}^6}\right)
  \right] M_*.
 \end{eqnarray}
 Notice that the term of $O(H^2/m_{3/2}^2)M_*$ is exactly equal to
$\bar{\phi}$. In the case where the potential of $X$ is logarithmic
like Eq.~(\ref{V(X)_approx}), formula for $\dot{H}$ is slightly
different:
 \begin{eqnarray}
  \dot{H} \sim - \frac{1}{1+L} H^2.
 \end{eqnarray}
 In this case, $\phi$ is given as
 \begin{eqnarray}
  \phi = \phi_{\rm osc} + \left[
  \frac{k_HH^2}{m_{3/2}^2}
  - \frac{36(5+2L-L^2)}{(1+L)^3} 
  \frac{k_HH^4}{m_{3/2}^4}
  + O \left(\frac{H^6}{m_{3/2}^6}\right)
  \right] M_*,
 \end{eqnarray}
 and the $O(H^2/m_{3/2}^2)M_*$ term agrees with $\bar{\phi}$ again. In
general, as far as $H\ll m_{3/2}$ and $\dot{H}\lesssim O(H^2)$, the
leading correction is always equal to $\bar{\phi}$. This is because,
in Eq.~(\ref{eq4phi_XH}), first two terms become of
$O(H^4/m_{3/2}^4)M_*$, and hence $\bar{\phi}$ has to be cancelled out
by $O(H^2/m_{3/2}^2)M_*$ term in $\phi$. As a result, the deviation
from the shifted minimum is always of $O(H^4/m_{3/2}^4)M_*$.

Since $\bar{\phi}$ smoothly goes to 0 as the Universe expands,
$\bar{\phi}$ term in $\phi$ is harmless. In other words, in the early
stage, $\phi$ oscillates around the shifted minimum $\bar{\phi}$, but
this minimum approaches to the true minimum as $H\rightarrow 0$.
Therefore, we just have to consider the deviation from
$\phi=\bar{\phi}$.

In our situation, potential of $X$ changes its behavior at $|X|\sim
v$; it is logarithmic for $|X|\gtrsim v$, and parabolic potential is
relevant once $X$ is trapped in the SUSY breaking minimum. Thus, the
solution to Eq.~(\ref{eq4phi_XH}) changes its behavior at $|X|\sim v$.
For example, if we match two cases at $L=1$ (though the matching point
is quite uncertain), $\phi_{\rm osc}$ is shifted as
 \begin{eqnarray}
  \phi_{\rm osc} \rightarrow
  \phi_{\rm osc} -\frac{45}{2} \frac{k_HH^4}{m_{3/2}^4} M_*.
 \label{phi_shift}
 \end{eqnarray}
 Notice that this shift is not $O(H^2/m_{3/2}^2)M_*$, but
$O(H^4/m_{3/2}^4)M_*$. In general, shift of this order may be
possible, especially when the potential changes its behavior. However,
shift cannot be $O(H^2/m_{3/2}^2)M_*$.

If this shift is larger than the averaged amplitude of $\phi_{\rm
osc}$, our argument may break down. Therefore, we require
 \begin{eqnarray}
  \phi_{\rm osc} \gtrsim \frac{H^4}{m_{3/2}^4} M_*.
 \end{eqnarray}
 When $|X|\gtrsim v$, $\phi_{\rm osc}$ is proportional to $|X|^{1/2}$
while $H$ depends on $|X|$ only logarithmically. On the other hand,
once $X$ is trapped in the SUSY breaking minimum, $\phi_{\rm osc}$ and
$H$ are both proportional to $(|X|-v)$. Therefore, this constraint
becomes most stringent when $|X|\sim v$. Since the expansion rate is
much smaller than the gravitino mass (see Eq.~(\ref{H_log})), above
constraint is very weak. Numerically, this requires $v\gtrsim 100{\rm
~GeV}$, even for $\zeta_2\sim 1$. Of course, if we adopt smaller value
of $\zeta_2$, lower bound on $v$ becomes less severe. Therefore, in
our case, this constraint is weak enough, and Hubble-induced term does
not change our argument.

\section{Evolution of $n_R$}
 \label{app:nR}
 \setcounter{equation}{0}

In this Appendix, we discuss how the $R$-number density evolves with
time. In particular, since $R$-symmetry breaking terms exist in the
potential, we need to know the evolution of $n_R$ with them.

We consider a case where $X$ oscillates with a large amplitude. In
this case, frequency of the oscillation is roughly given by $\sim
m_{\rm eff}$, where the effective mass $m_{\rm eff}$ is given by (see
Appendix~\ref{app:osc})
 \begin{eqnarray}
  m_{\rm eff}^2 = \frac{1}{2}
  \left( \frac{1}{X^*} \frac{\partial V}{\partial X} + 
  \frac{1}{X} \frac{\partial V}{\partial X^*} \right).
 \end{eqnarray}
 Notice that $m_{\rm eff}\gg H$ when $X$ is oscillating. Therefore, we
consider the time scale $m_{\rm eff}^{-1}\ll\delta t\ll H^{-1}$, for
which we neglect the change of $|X|$ and $H$. With the expansion of
the Universe, the amplitude of $X$ decreases as
 \begin{eqnarray}
  |X|R^p = {\rm const.},
 \label{XR^p=const}
 \end{eqnarray}
 with $p$ being a positive $O(1)$ constant. (For example, $p=3/2$ for
parabolic potential, and $p=3$ for logarithmic one.) Therefore, we
approximate the motion of $X$ as
 \begin{eqnarray}
  X \sim |X_0| e^{(im_{\rm eff}-pH)t}.
 \label{X-amp}
 \end{eqnarray}

With the $R$-symmetry breaking potential 
 \begin{eqnarray}
  V_{\not{R}} \sim -\frac{F_X^2}{M_*}
  (X^*+X) \times f(X^*X/M_*^2),
 \label{ap:V_Rbrk}
 \end{eqnarray}
 with $f(x)=k_0+k_1x+\cdots$, equation for the evolution of
$n_R=i(X^*\dot{X}-\dot{X}^*X)$ is given by
 \begin{eqnarray}
  \dot{n}_R + 3Hn_R = 
  2F_X^2{\rm Im} (X/M_*) \times f(|X|^2/M_*^2).
 \label{ap:dotn_r}
 \end{eqnarray}
 For the following discussion, it is more convenient to consider the
$R$-number in a comoving volume. For this quantity, the above equation
leads to
 \begin{eqnarray}
  \frac{d(R^3n_R)}{dt} = 
  R^3 \times 2F_X^2{\rm Im} (X/M_*) \times f(|X|^2/M_*^2).
 \label{dot_R3nR}
 \end{eqnarray}

We need to solve the above equation to obtain the resultant
$R$-number. For this purpose, we first take the average of the
right-hand side of this equation for the time scale $m_{\rm
eff}^{-1}\ll\delta t\ll H^{-1}$. For this time scale, we approximate
$|X|$ and $H$ to be (almost) constant. On the other hand, the change
of ${\rm Im}X$ is extremely non-adiabatic. In the flat background
(i.e., if $H=0$), average of ${\rm Im}X$ is supposed to vanish since
$X$ is in a periodic motion. However, in the actual situation, $H$ is
non-vanishing. By using Eq.~(\ref{X-amp}), average of ${\rm Im}X$ is
estimated as
 \begin{eqnarray}
  \langle {\rm Im}X \rangle \sim |X| \frac{H}{m_{\rm eff}}
  \sim \frac{1}{m_{\rm eff}} \frac{d|X|}{dt},
 \label{<ImX>}
 \end{eqnarray}
 where we related the expansion rate $H$ to $d|X|/dt$ by using
Eq.~(\ref{XR^p=const}).  In Eq.~(\ref{<ImX>}) and hereafter, we
neglect possible $O(1)$ coefficients since they do not change the
following argument.  Combining this equation with
Eq.~(\ref{dot_R3nR}), we obtain
 \begin{eqnarray}
  \frac{d(R^3n_R)}{d|X|} \sim
  R^3 \times \frac{2F_X^2}{M_*} \frac{f(|X|^2/M_*^2)}{m_{\rm eff}}.
 \end{eqnarray}
 By using the fact that $m_{\rm eff}|X|^2R^3$ is a constant of motion,
we integrate the above equation from $X=X_i$ to $X=X_f$ (with
$X_i>X_f$):
 \begin{eqnarray}
  R^3 n_R \sim
  m_{\rm eff}|X|^2R^3 \times \frac{2F_X^2}{M_*} 
  \times \sum_n \frac{k_n}{M_*^{2n}} \left(
  \frac{|X_f|^{2n-1}}{m_{\rm eff}^2(X_f)}
  - \frac{|X_i|^{2n-1}}{m_{\rm eff}^2(X_i)}
  \right).
 \label{nR_afterint}
 \end{eqnarray}
 If the second term in the parenthesis in Eq.~(\ref{nR_afterint}) wins
the first term, integration in large $|X|$ region is more important,
and major part of the $R$-number is generated when $X$ starts to move.
On the other hand, if the first term is dominant, we cannot neglect
the $R$-number generation in the later stage.

If the potential of $X$ is dominated by the logarithmic piece, $m_{\rm
eff}$ is proportional to $|X|^{-1}$. In this case, $R$-number
generation at large amplitude is more important for $n\geq 0$. As a
result, even if there is a linear $R$-symmetry breaking term in the
potential, $R$-number is generated when $X$ starts to move, and
$R$-symmetry is conserved with a good accuracy for a smaller value of
$|X|$. On the other hand, for parabolic potential, $m_{\rm eff}$ is a
constant. If the $R$-number violating potential is dominated by the
linear term ($n=0$), contribution at small $|X|$ becomes important.
However, in our scenario, we assume that the linear term is suppressed
enough when $X$ starts to move, and that the $R$-number violating
effect starts with cubic term ($n=1$). In this case, $R$-number
asymmetry is again generated when $X$ starts to move.

In the actual situation, $X$ starts to move with a quadratic
potential, and at some stage, logarithmic piece takes over. With the
assumption that the linear term is suppressed enough, $R$-number is
generated when $X$ starts to move, and afterwards, $R$-number in the
comoving volume is conserved well.

\section{Scalar Field in the Expanding Universe}
 \label{app:osc}
 \setcounter{equation}{0}

In this Appendix, we derive a convenient formula for the evolution of
the scalar field $\varphi$ in periodic motion.  For simplicity, we
consider the case where the amplitude of the scalar field is (almost)
constant in a time scale of the periodic motion and also the potential
for $\varphi$ depends only on $|\varphi|$.

From the virial theorem, we obtain
 \begin{eqnarray}
  2\langle {\cal K} \rangle =
  \left\langle \varphi \frac{\partial V}{\partial\varphi}
  + \varphi^* \frac{\partial V}{\partial\varphi^*} \right\rangle,
 \end{eqnarray}
 where ${\cal K}=\dot{\varphi}^*\dot{\varphi}$ is the kinetic energy
of $\varphi$, and the bracket represents the time average.

The field equation for $\varphi$ is given by
 \begin{eqnarray}
  \ddot{\varphi} + 3H\dot{\varphi} 
  + \frac{\partial V}{\partial\varphi^*}
  = 0.
 \end{eqnarray}
 Multiplying this equation by $\dot{\varphi}^*$, and using the
definition of ${\cal K}$, we obtain
 \begin{eqnarray}
  \dot{\cal K} + 6H{\cal K} + 
  \left( \dot{\varphi}^* \frac{\partial V}{\partial\varphi^*}
  + \dot{\varphi} \frac{\partial V}{\partial\varphi} \right) = 0.
 \end{eqnarray}

Now, we are at the position to consider the evolution of the scalar
field. For this purpose, we define
 \begin{eqnarray}
  S^2 = |\varphi|^2
  \left( \varphi \frac{\partial V}{\partial\varphi} + 
  \varphi^* \frac{\partial V}{\partial\varphi^*} \right),
 \end{eqnarray}
 and consider the evolution of $\langle S^2\rangle$. By taking the
derivative of $\langle S^2\rangle$ with respect to time, we obtain
 \begin{eqnarray}
  \frac{d\langle S^2\rangle}{dt} &=&
  \left\langle \frac{d|\varphi|^2}{dt}
  \left( \varphi \frac{\partial V}{\partial\varphi} + 
  \varphi^* \frac{\partial V}{\partial\varphi^*} \right)
  + |\varphi|^2 \frac{d}{dt}
  \left( \varphi \frac{\partial V}{\partial\varphi} + 
  \varphi^* \frac{\partial V}{\partial\varphi^*} \right) \right\rangle
 \nonumber \\ &=&
  \left\langle \frac{d|\varphi|^2}{dt}
  \left( \varphi \frac{\partial V}{\partial\varphi} + 
  \varphi^* \frac{\partial V}{\partial\varphi^*} \right)
  + |\varphi|^2 \dot{\cal K} \right\rangle
 \nonumber \\ &=&
  -12 H \left\langle |\varphi|^2 {\cal K} \right\rangle
 \nonumber \\ &=&
  -6H \langle S^2\rangle,
 \end{eqnarray}
 where we used the fact that the potential $V$ is a function of
$|\varphi|$. By solving the above equation, we obtain
 \begin{eqnarray}
  \langle S^2\rangle R^6 = {\rm const}.
 \end{eqnarray}
 The scalar field evolves by following this relation.

For a more intuitive understanding, it is convenient to define the
``effective mass'' from the potential $V$:
 \begin{eqnarray}
  m_{\rm eff}^2 = \frac{1}{2}
  \left( \frac{1}{\varphi^*} \frac{\partial V}{\partial\varphi} + 
  \frac{1}{\varphi} \frac{\partial V}{\partial\varphi^*} \right).
 \end{eqnarray}
 With this effective mass, evolution of the scalar field is given by
 \begin{eqnarray}
  m_{\rm eff} |\varphi|^2 R^3 = {\rm const}.
 \label{mphi2=const}
 \end{eqnarray}
 For example, in the case of parabolic potential
$V=m_\varphi^2|\varphi|^2$, $m_{\rm eff}$ does not depend on
$\varphi$, and hence $|\varphi|$ scales as $R^{-3/2}$, while
$|\varphi|\propto R^{-1}$ for quartic potential $V\propto|\varphi|^4$.

Notice that, in the flat space ($H=0$), 
 \begin{eqnarray}
  \varphi = \varphi_0 e^{\pm im_{\rm eff}t},
 \label{e^imt}
 \end{eqnarray}
 satisfies the equation of motion of $\varphi$ for any value of
$\varphi_0$, if the potential of $\varphi$ depends only on
$|\varphi|$.  (In Eq.~(\ref{e^imt}), $m_{\rm eff}$ is evaluated at
$\varphi =\varphi_0$.)  Therefore, $m_{\rm eff}$ can be understood as
a frequency of the periodic motion.

Since $m_{\rm eff} |\varphi|^2$ is proportional to the volume of the
phase space for the periodic motion, a physical interpretation of
Eq.~(\ref{mphi2=const}) is that the phase space volume in a comoving
volume is conserved as the Universe expands.


\newpage

\begin{thebibliography}{99}

\bibitem{GMSB_origs}
  M.~Dine and A.E.~Nelson,
    {\sl Phys.~Rev.} {\bf D48}, 1277 (1993);
  M.~Dine, A.E.~Nelson and Y.~Shirman,
    {\sl Phys.~Rev.} {\bf D51}, 1362 (1995);
  M.~Dine, A.E.~Nelson, Y.~Nir and Y.~Shirman, 
    {\sl Phys.~Rev.} {\bf D53}, 2658 (1996).

\bibitem{hph9801271}
  For review, see, for example, G.F.~Giudice and R.~Rattazzi,
    hep-ph/9801271.

\bibitem{PRL48-223}
  H.~Pagels and J.R.~Primack,
    {\sl Phys.~Rev.~Lett.} {\bf 48} (1982) 223.

\bibitem{PLB303-289}
  T.~Moroi, H.~Murayama and M.~Yamaguchi,
    {\sl Phys.~Lett.} {\bf B303} (1993) 289.

\bibitem{PRD56-1281}
  A.~de Gouv\^ea, T.~Moroi and H.~Murayama,
    {\sl Phys.~Rev.} {\bf D56} (1997) 1281.

\bibitem{thermal_inf}
  D.H. Lyth and E.D. Stewart, 
    {\sl Phys.~Rev.~Lett.} {\bf 75} (1995) 201;
    {\sl Phys.~Rev.} {\bf D53} (1996) 1784.

\bibitem{PRL79-18}
  H.~Murayama,
    {\sl Phys.~Rev.~Lett.} {\bf 79} (1997) 18.

\bibitem{NPB510-12}
  S.~Dimopoulos, G.~Dvali, R.~Rattazzi and G.~Giudice,
    {\sl Nucl.~Phys.} {\bf B510} (1998) 12.

\bibitem{PLB413-336}
  S.~Dimopoulos, G.~Dvali and R.~Rattazzi,
    {\sl Phys.~Lett.} {\bf B413} (1997) 336.

\bibitem{hph9804450}
  K.~Agashe,
    hep-ph/9804450.

\bibitem{nonpert}
  K.-I.~Izawa, Y.~Nomura, K.~Tobe and T. Yanagida,
    {\sl Phys.~Rev.} {\bf D56} (1997) 2886;
  Y.~Nomura and K.~Tobe,
    hep-ph/9708377 .

\bibitem{otherDGM}
  E.~Poppitz and S.P.~Trivedi, 
    {\sl Phys.~Rev.} {\bf D55} (1997) 5508;
  N.~Arkani-Hamed, H.~Murayama and J.~March-Russell, 
    {\sl Nucl.~Phys.} {\bf B509} (1998) 3;
  M.A.~Luty,
    {\sl Phys.~Lett.} {\bf B414} (1997) 71;
  Y.~Shirman,
    {\sl Phys.~Lett.} {\bf B417} (1998) 281;
  M.A.~Luty and J.~Terning,
    {\sl Phys.~Rev.} {\bf D57} (1998) 6799;
  N.~Arkani-Hamed, M.A.~Luty and J.~Terning,
    hep-ph/9712389.

\bibitem{PLB105-267}
  E.~Witten,
    {\sl Phys.~Lett.} {\bf B105} (1981) 267.

\bibitem{IzaYanInt}
  K.-I.~Izawa and T.~Yanagida,
    {\sl Prog.~Theor.~Phys.} {\bf 95} (1996) 829;
  K.~Intriligator and S.~Thomas,
    {\sl Nucl.~Phys.} {\bf B473} (1996) 121.

\bibitem{DinRanTho}
  M.~Dine, L.~Randall and S.~Thomas,
    {\sl Phys.~Rev.~Lett.} {\bf 75} (1995) 398;
    {\sl Nucl.~Phys.} {\bf B458} (1996) 291.

\bibitem{NPB249-361}
  I.~Affleck and M.~Dine,
    {\sl Nucl.~Phys.} {\bf B249} (1985) 361.

\bibitem{PRep267-195}
  See, for example, G.~Jungman, M.~Kamionkowski and K.~Griest,
    {\sl Phys.~Rep.} {\bf 267} (1996) 195.

\bibitem{hph9712515}
  L.J.~Hall, T.~Moroi and H.~Murayama,
    {\sl Phys.~Lett.} {\bf B424} (1998) 305.

\bibitem{PLB386-189}
  S.~Borgani, A.~Masiero and M.~Yamaguchi,
    {\sl Phys.~Lett.} {\bf B386} (1996) 189.

\bibitem{messenger-DM}
  S.~Dimopoulos, G.F.~Giudice and A.~Pomarol,
    {\sl Phys.~Lett.} {\bf B389} (1996) 37;
  T.~Han and R.~Hempfling,
    {\sl Phys.~Lett.} {\bf B415} (1997) 161.

\bibitem{QCD_br}
  N.~Arkani-Hamed, C.D. Carone, L.J. Hall and H.~Murayama,
    {\sl Phys.~Rev.} {\bf D54} (1996) 3072;
  I.~Dasgupta, B.A.~Dobrescu and L.~Randall,
    {\sl Nucl.~Phys.} {\bf B483} (1997) 95.

\bibitem{PLB113-231}
  G.~Veneziano and S.~Yankielowicz,
    {\sl Phys.~Lett.} {\bf B113} (1982) 231.

\bibitem{NDA}
  M.A.~Luty,
    {\sl Phys.~Rev.} {\bf D57} (1998) 1531;
  A.G.~Cohen, D.B.~Kaplan and A.E. Nelson,
    {\sl Phys.~Lett.} {\bf B412} (1997) 301.

\bibitem{NPB426-3}
  J.~Bagger, E.~Poppitz and L.~Randall,
    {\sl Nucl.~Phys.} {\bf B426} (1994) 3.

\bibitem{PLB129-177}
  A.~Linde,
    {\sl Phys.~Lett.} {\bf B129} (1983) 177.

\bibitem{hph9805300}
  B.A.~Campbell, M.K.~Gaillard, H.~Murayama and K.A.~Olive,
    hep-ph/9805300.

\bibitem{PRD56-7597}
  M.~Kasuya and M.~Kawasaki,
    {\sl Phys.~Rev.} {\bf D56} (1997) 7597.

\bibitem{PRD50-4821}
  M.~Nagasawa and M.~Kawasaki,
    {\sl Phys.~Rev.} {\bf D50} (1994) 4821.

\bibitem{hph9805405}
  See, for example, E.~Holtmann, M.~Kawasaki, K.~Kohri and T.~Moroi,
    hep-ph/9805405.

\bibitem{PLB328-354}
  S.~Davidson, H.~Murayama and K.A. Olive,
    {\sl Phys.~Lett.} {\bf B328} (1994) 354.

\bibitem{HBstar}
  G.G.~Raffelt,
    {\sl Stars as Laboratories for Fundamental Physics: The
Astrophysics of Neutrinos, Axions, and Other Weakly Interacting
Particles} (University of Chicago Press, 1996).

\bibitem{UVlight}
  M.S.~Turner, 
    {\sl Phys.~Rev.~Lett.} {\bf 59} (1987) 2489;
  M.T.~Ressell,
    {\sl Phys.~Rev.} {\bf D44} (1991) 3001;
  J.M.~Overduin and P.S.~Wesson,
    {\sl Astrophys.~J.} {\bf 414} (1993) 449.

\bibitem{SN1987A}
  M.S.~Turner,
    {\sl Phys.~Rev.~Lett.} {\bf 60} (1988) 1797;
  A.~Burrows, M.S.~Turner and R.P.~Brinkmann,
    {\sl Phys.~Rev.} {\bf D39} (1989) 1020;
  A.~Burrows, M.T.~Ressell and M.S.~Turner,
    {\sl Phys.~Rev.} {\bf D42} (1990) 3297.

\bibitem{PRL65-960}
  J.~Engel, D.~Seckel and A.C.~Hayes,
    {\sl Phys.~Rev.~Lett.} {\bf 65} (1990) 960.

\bibitem{PRL79-2632}
  M.~Dine and A.~Riotto,
    {\sl Phys.~Rev.~Lett.} {\bf 79} (1997) 2632;
  S.~Dimopoulos, G.~Dvali and R.~Rattazzi,
    {\sl Phys.~Lett.} {\bf B410} (1997) 119.

\bibitem{Xray}
  M.~Kawasaki and T.~Yanagida,
    {\sl Phys.~Lett.} {\bf B399} (1997) 45;
  T.~Asaka, J.~Hashiba, M.~Kawasaki and T.~Yanagida,
    hep-ph/9802271.

\bibitem{NPB416-46}
  A.E.~Nelson and N.~Seiberg,
    {\sl Nucl.~Phys.} {\bf B416} (1994) 46.

\bibitem{NPB207-96}
  L.~Alvarez-Gaum\'{e}, M.~Claudson and M.~Wise, 
    {\sl Nucl.~Phys.} {\bf B207} (1982) 96.

\end{thebibliography}
\end{document}